\documentclass[%
reprint,
superscriptaddress,
nobibnotes,
amsmath,amssymb,
aps,
pra,
floatfix,
]{revtex4-2}
\usepackage{packages}

\begin{document}
\title{Quantum Averaging Theory for Multi-Timescale Driven Quantum Systems}% on Multiple Timescales}
\author{Kristian D. Barajas}

% Dear Curious Physicist,
%
% Congratulations! You have unlocked the hidden message. Whether you stumbled upon this by accident or have an insatiable curiosity for LaTeX source files, I salute your dedication.
%
% Thank you for taking the time to explore my first paper--I genuinely appreciate your interest in what I’ve put together here. If you've made it this far, I can only assume you have a keen eye for detail and a love for the intricacies of the Magnus expansion (or maybe you’re just debugging my references, in which case, good luck).
%
% A little bit about my PhD journey: I was an APS Bridge student from 2018–2019 (and very proud of this!) and then transitioned into my doctoral program at UCLA. I was guided by a wonderful theorist and, more importantly, hilarious and good-natured person, Dr. Alex Levine, who unfortunately passed away in 2022. Since then, I was adopted by an experimental group and shifted to the theory of quantum control under Dr. Wes Campbell, where I now spend my days thinking about driven quantum systems and multi-timescale dynamics. Outside of research, I enjoy board games, traveling, and running a D&D campaign for my close friends. (If you were hoping for quantum-themed D&D rules, you may be disappointed—but perhaps inspired to homebrew some!)
%
% If you ever want to chat about physics, gaming, or anything in between, feel free to reach out (kdbarajas@live.com). In the meantime, enjoy digging through my LaTeX spaghetti code!
%
% Best wishes and happy compiling,
% Kristian D. Barajas

\affiliation{%
 Department of Physics \& Astronomy, University of California -- Los Angeles, California 90095, USA
}

\affiliation{%
Mani L. Bhaumik Institute for Theoretical Physics %, \\
% UCLA Department of Physics and Astronomy 
 \\ University of California -- Los Angeles, Los Angeles, California 90095, USA 
}

\affiliation{%
 UCLA Center for Quantum Science and Engineering (CQSE)
 \\ University of California -- Los Angeles, Los Angeles, California 90095, USA
}

\affiliation{
 NSF Challenge Institute for Quantum Computation
 \\ University of California -- Berkeley, Berkeley, California 94720, USA}

\author{Wesley C. Campbell}%
%\email{wes@physics.ucla.edu}
\affiliation{%
 Department of Physics \& Astronomy, University of California -- Los Angeles, California 90095, USA
}

\affiliation{%
 UCLA Center for Quantum Science and Engineering (CQSE)
 \\ University of California -- Los Angeles, Los Angeles, California 90095, USA
}

\affiliation{
 NSF Challenge Institute for Quantum Computation
 \\ University of California -- Berkeley, Berkeley, California 94720, USA}

\date{March 1, 2025}
\vspace{1cm}

\begin{abstract}
We present a multi-timescale Quantum Averaging Theory (QAT), a unitarity-preserving generalized Floquet framework for analytically modeling periodically and almost-periodically driven quantum systems across multiple timescales. 
By integrating the Magnus expansion with the method of averaging on multiple scales, QAT captures the effects of both far-detuned and near-resonant interactions on system dynamics.
The framework yields an effective Hamiltonian description while retaining fast oscillatory effects within a separate dynamical phase operator, ensuring accuracy across a wide range of driving regimes. 
We demonstrate the rapid convergence of QAT results toward exact numerical solutions in both detuning regimes for touchstone problems in quantum information science.

\end{abstract}
\maketitle

% \section{Introduction}\label{sec:intro}
Driven quantum systems, characterized by interactions with time-dependent external fields, are central to both fundamental physics and advancements in quantum technologies. 
These systems often exhibit multi-timescale dynamics, where slowly-varying interactions are modulated by fast oscillatory effects, challenging standard analytic and numerical modeling techniques.
In the treatment of simple systems where long-time dynamics are easily distinguishable from transient effects, fast timescale interactions are often averaged out. 
However, capturing how these interactions modulate slower effective dynamics in complex physical systems requires careful modeling to produce accurate analytic results.

The construction of effective Hamiltonians that capture the slowly-varying dynamics has emerged as a powerful tool for simplifying complex dynamics and providing intuition for the long-time behavior. 
Traditional methods, such as the rotating-wave approximation (RWA) \cite{footAtomicPhysics2005,sakuraiModernQuantumMechanics2020} and adiabatic elimination \cite{footAtomicPhysics2005,bergmannPerspectiveStimulatedRaman2015,sanzAdiabaticEliminationEffective2016}, rely on heuristic arguments, limiting their potential for systematic improvements in accuracy. 
More-rigorous approaches, such as the Schrieffer–Wolff expansion \cite{schriefferRelationAndersonKondo1966} and projector-based techniques \cite{paulischAdiabaticEliminationHierarchy2014,sanzAdiabaticEliminationEffective2016}, address these limitations by isolating relevant subspaces while suppressing non-resonant interactions. 
James \textit{et al.} \cite{jamesEffectiveHamiltonianTheory2007} proposed a low-pass filter technique to eliminate high-frequency contributions in a Dyson expansion, providing an approximately Hermitian framework. 
While insightful, these methods often struggle to capture the full influence of fast-varying interactions on long-term dynamics, leaving critical gaps in understanding.
Quantum resonance theory, as outlined by Frasca \cite{frascaUnifiedTheoryQuantum1992,frascaTheoryQuantumResonance1998}, offers a robust foundation leveraging multi-timescale perturbative analysis. 
However, it lacks generality and does not produce an effective Hamiltonian, limiting its practical applicability.
Furthermore, these methods share a critical drawback: they are only approximately unitary for sufficiently small perturbations, leading to non-unitary artifacts in the results \cite{blanesMagnusExpansionIts2009}.
Further, their scalability to higher-order corrections is limited, which can make the process of achieving systematic improvements in accuracy both tedious and impractical.

In contrast, the Magnus expansion preserves unitarity \cite{magnusExponentialSolutionDifferential1954,blanesMagnusExpansionIts2009} and offers a versatile Lie-algebraic framework for generating effective Hamiltonians utilized in Floquet-Magnus theory \cite{casasFloquetTheoryExponential2001,kuwaharaFloquetMagnusTheory2016}, average Hamiltonian theory \cite{brinkmannIntroductionAverageHamiltonian2016}, and Van Vleck perturbation theory \cite{rahavEffectiveHamiltoniansPeriodically2003,goldmanPeriodicallyDrivenQuantum2014,eckardtHighfrequencyApproximationPeriodically2015,bukovUniversalHighFrequencyBehavior2015}.
These high-frequency expansions are applied to great effect in the field of Floquet engineering \cite{hemmerichEffectiveTimeindependentDescription2010,goldmanPeriodicallyDrivenQuantum2014,eckardtHighfrequencyApproximationPeriodically2015} to tailor time-periodic interactions to simulate and probe quantum systems with novel properties.
This has led to the experimental realization of dynamical localization \cite{eckardtExploringDynamicLocalization2009}, synthetic magnetic fields \cite{roushanChiralGroundstateCurrents2017,aidelsburgerExperimentalRealizationStrong2011,andersonMagneticallyGeneratedSpinOrbit2013}, topological systems \cite{jotzuExperimentalRealizationTopological2014,winterspergerRealizationAnomalousFloquet2020}, and to the anomalous and fractional quantum Hall effects \cite{sorensenFractionalQuantumHall2005,mciverLightinducedAnomalousHall2020,leonardRealizationFractionalQuantum2023}.
However, these high-frequency expansions are typically limited to periodic systems and falter when applied to incommensurate frequencies or multi-timescale phenomena. 
Consequently, addressing the unitary dynamics of driven systems across disparate timescales remains an open problem.

To resolve these challenges, we present a quantum averaging theory (QAT) framework that integrates the unitarity-preserving Magnus expansion with the method of averaging on multiple timescales.
The method of averaging (MA), originally developed in nineteenth-century celestial mechanics \cite{poincareMethodesNouvellesMecanique1892} to study nonlinear dynamical systems, identifies time-averaged dynamics by systematically eliminating high-frequency perturbations \cite{kamelPerturbationMethodTheory1970,caryLieTransformPerturbation1981}.
Its extension to Hilbert spaces, pioneered by Buitelaar \cite{buitelaarMethodAveragingBanach1993} and Scherer \cite{schererNewPerturbationAlgorithms1997}, provided the foundation for applying averaging techniques to quantum systems, forming the basis of our approach.
For systems that evolve with both short- and long-time effects, a multi-timescale analysis (MTSA) is employed to regularize the dynamics and enable a valid long-time expansion \cite{nayfehMethodMultipleScales2000}.
Building on these advancements, quantum averaging theory unifies Magnus-based high-frequency expansions, systematizes multi-timescale analysis, and establishes a cohesive, unitarity-preserving framework for theoretical and experimental progress.

Quantum averaging theory addresses multi-timescale dynamics by factorizing the unitary evolution into a fast and slow propagator.
The slow propagator is governed by a Hermitian effective Hamiltonian derived from an effective (i.e., renormalization group) equation, while the fast oscillatory effects are retained in a separate dynamical phase operator.
This phase operator ensures unitarity and plays a role analogous to the micromotion operator in Floquet-Magnus theory \cite{casasFloquetTheoryExponential2001}.
QAT operates in two distinct regimes: far-detuned interactions and multi-timescale interactions, determined by the relative strength of the perturbation amplitude to the drive frequencies.
In the far-detuned regime, the method of averaging generalizes the principles of well-known Magnus-based time-periodic expansions \cite{rahavEffectiveHamiltoniansPeriodically2003,casasFloquetTheoryExponential2001,goldmanPeriodicallyDrivenQuantum2014,eckardtHighfrequencyApproximationPeriodically2015} to systems with almost-periodic, multi-modal drives \cite{casasTimedependentPerturbationTheory2015,vinogradovPhaseModulatedLee2001}.
In the multi-timescale regime, QAT reproduces first-order approximations from established techniques, such as the rotating-wave approximation and adiabatic elimination, but systematically extends beyond them, enabling higher-order corrections in a unitary and analytically controlled manner.
Although the examples explored in this work are motivated by quantum optics, the QAT framework is generally applicable to any driven quantum system with multiple timescales, including systems in condensed matter, atomic physics, or engineered quantum platforms.

\section{The Quantum Perturbation Problem}\label{sec:perturbation_problem}

Consider a weakly-perturbed and unitary (i.e. reversible) quantum system described by the state vector 
\begin{equation}\label{eq:qat:state_vector}
\ket{\psi(s,s_0)} = \hat{U}_\mathrm{S}(s,s_0) \ket{\psi(s_0)}
\end{equation}
or, more generally, by the density matrix
\begin{equation}
\hat{\rho}(s,s_0) = \hat{U}_\mathrm{S}(s,s_0) \hat{\rho}(s_0) \hat{U}^\dagger_\mathrm{S}(s,s_0)
\end{equation}
where $s \propto t$ is a scaled, dimensionless time variable. 
The system evolves under the unitary time-evolution operator (or \textit{propagator}) in the Schr\"odinger picture $\hat{U}_\mathrm{S}(s,s_0)$ governed by the total Hamiltonian
\begin{equation}\label{eq:qat:SchroHamiltonian}
    \hat{H}_\mathrm{S}(s;\lambda) = \hat{H}_0(s) + \hat{V}(s;\lambda)
\end{equation}
where $\hat{H}_0(s)$ is a solvable, ``unperturbed'' Hamiltonian and $\hat{V}(s;\lambda) = \sum_{n=1}^{\infty} \lambda ^{n} \hat{V}^{(n)}(s)$ is a bounded perturbation with $0 \le \lambda < 1$ \cite{sakuraiModernQuantumMechanics2020}.
The choice of unperturbed Hamiltonian is flexible and typically field-dependent.

We assume $s$ to be the time $t$ scaled by the characteristic frequency of the unperturbed Hamiltonian, $\omega_0 \sim \lVert \hat{H}_0(t) \rVert/\hbar$, where the relevant energy scale for arbitrary $\hat{H}_0$ is given by its spectral norm $\lVert \hat{H}_0 \rVert$ (\textit{i.e.}, the largest singular value).  
For example, in a two-level system (2LS) with $\hat{H}_0(t)/\hbar = \omega_{eg} \hat{\sigma}_z/2$, one could choose $s = \omega_{eg}t$ resulting in $\hat{H}_\mathrm{S}(s) \doteq \hat{H}_\mathrm{S}(t = s /\omega_{eg})/\hbar \omega_{eg}$ with $\hat{H}_0(s) = \hat{\sigma}_z/2$ and $\lambda \sim \lVert \hat{V}(t) \rVert / \hbar \omega_{eg}$. 

We will find it useful to work instead with the $s_0$-invariant propagator solution $\hat{U}_\mathrm{S}(s)$ related to the time-centered propagator in \eqref{eq:qat:state_vector} by
\begin{equation}
   \hat{U}_\mathrm{S}(s,s_0) = \hat{U}_\mathrm{S}(s) \hat{U}^{\dagger}_\mathrm{S}(s_0).
\end{equation}
where $\hat{U}_\mathrm{S}(s_0,s_0)= \hat{U}_\mathrm{S}(s_0) \hat{U}^{\dagger}_{\mathrm{S}}(s_0) = \idty$.
We require that as $\lambda \to 0$, the propagator reduces to $\hat{U}_0(s)$ governed by $\hat{H}_0(s)$.
Following standard perturbative treatment \cite{sakuraiModernQuantumMechanics2020} we factorize the total propagator as \footnote{The factorized solution from the variation of constants is guaranteed by the condition that the propagator $\hat{U}_0(s)$ be uniformly bounded \cite{buitelaarMethodAveragingBanach1993}.}
\begin{equation}\label{eq:qat:total_propagator}
	\begin{split}
		\hat{U}_\mathrm{S}(s;\lambda) &= \hat{U}_0(s) \, \hat{U}_I(s;\lambda) 
	\end{split}
\end{equation}
where the interaction propagator $\hat{U}_I(s;\lambda)$ is governed by the interaction Hamiltonian
\begin{subequations}\label{eq:qat:InteractionHamiltonian}
\begin{equation}
	\hat{H}_I(s;\lambda) = \sum_{n=1}^{\infty} \lambda^n \hat{H}^{(n)}_I(s) \\ 
     \hat{H}^{(n)}_I(s) = \hat{U}_0^\dagger(s) \hat{V}^{(n)}(s) \hat{U}_0(s)
\end{equation}
\end{subequations}
and satisfies the interaction-picture Schr\"odinger equation
\begin{equation} \label{eq:qat:interaction_eq}
	i \, \partial_s \hat{U}_I(s;\lambda) = \hat{H}_I(s;\lambda)\, \hat{U}_I(s;\lambda),
\end{equation}
which remains to be solved.
In the following treatment, we assume that $\hat{H}_I(s)$ is reasonably well-behaved as to be expanded in an almost-periodic Fourier series.
We define the ``frequency vector''  $\vec{\Lambda} = (\Lambda_{\omega_1}, \ldots,\Lambda_{\omega_q}) > 0$ from the set of unique Fourier frequencies of $\hat{H}_I(s)$ such that $\hat{H}_I(s;\lambda) \equiv  \hat{\mathcal{H}}_I(\vec{\Lambda}s;\lambda)$, which is $2\pi$-periodic with respect to each Fourier generalized phase $\Lambda_k s$ for $\Lambda_k \in \vec{\Lambda}$.
Without loss of generality, the interaction Hamiltonian may be expanded in terms of the Fourier modes as
\begin{equation}\label{eq:qat:Fourier_basis}
	\hat{H}_I ^{(n)}(s) = \hat{H}_{I,0}^{(n)} +
  \sum_{\mathclap{\Lambda_{\omega_k} \in \vec{\Lambda}}} \left(  \hat{h} ^{(n)}_{I,k} \, e^{-i s \Lambda_{\omega_k}} + h.c. \right)
\end{equation}
where $\hat{h}_{I,k}^{(n)}$ is the mode operator associated with the (dimensionless) Fourier base frequency $\Lambda_{\omega_{k}} \doteq {\omega_{k}}/{\omega_0} > 0$.
Assuming all Fourier modes interact, the discrete frequency spectrum generated by $\hat{H}_I(s;\lambda)$ contains the base frequencies and their sum and difference combinations.

To approximate the dynamics one may turn to standard time-dependent perturbation theory, seeking a solution  through a Dyson series expansion in $\lambda$,
\begin{equation}\label{eq:qat:Dyson_series}
    \hat{U}_I(s;\lambda) = \idty + \sum_{n=1}^{\infty} \lambda^n \, \hat{U}^{(n)}_I(s),
\end{equation}
or with the increasingly popular Magnus expansion \cite{magnusExponentialSolutionDifferential1954}
\begin{equation}
    \hat{U}_I(s;\lambda) = e^{-i \hat{\Phi}_I(s;\lambda)}, \quad \hat{\Phi}_I(s;\lambda) = \sum_{n = 1}^\infty \lambda^n \hat{\Phi}_I^{(n)}(s).
\end{equation}
depending on a (Hermitian) dynamical phase operator $\hat{\Phi}_I(s;\lambda)$ as detailed in Ref.~\cite{blanesMagnusExpansionIts2009} and references therein.
The algebraic Dyson expansion has been shown to have a number of qualitative drawbacks for describing quantum dynamics \cite{blanesMagnusExpansionIts2009, casasFloquetTheoryExponential2001,eckardtHighfrequencyApproximationPeriodically2015}; chiefly, the transformation is only unitary in the infinite resummation of the asymptotic series.
Therefore, truncated at some finite order, one yields an asymptotic approximation, with unitarity only preserved in the Magnus approach.
For these reasons, the Magnus expansion will serve as the foundation of our perturbative analysis.

In any case, whether unitarity is preserved or not, regular perturbation theory is generally only valid at short-times and when the system is driven far from resonance.
To illustrate this point, consider the first-order Magnus expansion result:
\begin{equation}\label{eq:qat:PT_first_order}
\begin{aligned}
    \hat{\Phi}^{(1)}_I(s) ={}& \int_{}^s d{s'} \: \hat{H}^{(1)}_I(s') \\
    ={}& \hat{H}_{I,0}^{(1)} s + \sum_{\Lambda_{\omega_k} \in\vec{\Lambda}}\left( \frac{e^{-i s \Lambda_{\omega_{k}}}}{-i\Lambda_{\omega_{k}}} \hat{h}_{I,k}^{(1)} + h.c.\right).
\end{aligned}
\end{equation}
We identify two concerns that limit the asymptotic validity for modeling the long-time behavior of driven quantum systems.
The first is the secular term $\hat{H}^{(1)}_{I,0} s$ that diverges as $s \rightarrow \infty$, which leads to an asymptotic breakdown when $\lambda s \ge 1$.
This term limits the accuracy of the approximation to a short duration $s\ll \mathcal{O}(1/\lambda)$.
The second is from the time-harmonic terms where any near-resonant frequency $\Lambda_{\omega_{k}} \le \lambda$ produces the infamous small denominator problem which breaks asymptotic validity \cite{sandersAveragingMethodsNonlinear2007}.
Even when validity holds at first order, the same concerns repeat at the next and so on.

To address these issues we apply the \textit{method of averaging} (MA) on Hilbert spaces, a non-secular perturbative approach used in the study of non-linear dynamical systems over long times \cite{buitelaarMethodAveragingBanach1993,schererNewPerturbationAlgorithms1997,sandersAveragingMethodsNonlinear2007}.
Rather than directly expanding $\hat{U}_I(s;\lambda)$, the basic idea is that the relevant, long-time dynamics are adequately described by an effective propagator $\hat{U}_{I,\mathrm{eff}}(s;\lambda)$, which satisfies the time-averaged, effective interaction picture equation
\begin{empheq}[box=\tightfbox]{equation}\label{eq:qat:QAT_eff_system}
i \frac{d}{ds}\hat{U}_{I,\mathrm{eff}}(s;\lambda) = \hat{H}_{I,\mathrm{eff}}(s;\lambda)\hat{U}_{I,\mathrm{eff}}(s;\lambda)
\end{empheq}
governed by a coarse-grained effective Hamiltonian
\begin{equation}
    \hat{H}_{I,\mathrm{eff}}(s;\lambda) = \sum_{k=1}^{\infty} \lambda ^{k} \, \hat{H}_{I,\mathrm{eff}}^{(k)}(s).
\end{equation}
The effective Hamiltonian, unknown \textit{a priori}, is expected to describe the slowly-varying $s \gtrsim \mathcal{O}(1/\lambda)$ dynamics of $\hat{H}_I(s;\lambda)$, insensitive to fast-varying details.
To see this more clearly, if we introduce another rescaled time variable $\tau = \lambda s$ the effective equation becomes
\begin{equation}\label{eq:qat:QAT_eff_system_slowtime}
i \frac{d}{d\tau}\hat{U}_{I,\mathrm{eff}}(\tau) = \left(\hat{H}_{I,\mathrm{eff}}^{(1)}(\tau) + \hat{H}_{I,\mathrm{eff}}^\prime(\tau;\lambda) \right) \, \hat{U}_{I,\mathrm{eff}}(\tau
)
\end{equation}
where $\hat{H}_{I,\mathrm{eff}}^\prime(\tau;\lambda) = \sum_{k\ge1} \lambda^k \hat{H}_{I,\mathrm{eff}}^{(k+1)}(\tau)$, yielding a perturbation problem for the slow $\tau$-time dynamics.
The exact (or perturbative) solution to the effective equation is valid at least on times $0 \le \tau \le 1$ \cite{sandersAveragingMethodsNonlinear2007}, surpassing the expected short-time $s \ll \mathcal{O}(1/\lambda)$ accuracy from regular perturbation theory as desired.
Moreover, due to its slowly-varying time-dependence, we will find that solving the effective equation for $\hat{U}_\mathrm{eff}(s;\lambda)$ is often simpler, both analytically and numerically, than eq.~\eqref{eq:qat:interaction_eq}.

To maintain an exact expansion, we seek a unitary transformation between \eqref{eq:qat:interaction_eq} and \eqref{eq:qat:QAT_eff_system} in the QAT factorized form
\begin{empheq}[box=\tightfbox]{equation}\label{eq:qat:interaction_propagator_factorized}
		\hat{U}_I(s;\lambda) = \hat{U}_\mathrm{fast}(s;\lambda) \, \hat{U}_\mathrm{eff}(s;\lambda)
\end{empheq}
where the fast-varying propagator $\hat{U}_\mathrm{fast}(s;\lambda)$ modulates the slowly-varying envelope generated by $\hat{U}_\mathrm{eff}(s;\lambda)$.
Hence, the aim is to iteratively construct an effective Hamiltonian that: (1) regularizes the perturbative expansion of $\hat{U}_{I}(s;\lambda)$, and (2) provides a useful description of the long-time dynamics.
The process will ensure that \eqref{eq:qat:interaction_propagator_factorized} embodies the full interaction picture dynamics up to some finite order. 

Before proceeding we provide a brief overview of the main result of this paper.
For a multi-modal interaction Hamiltonian described by eq.~\eqref{eq:qat:InteractionHamiltonian}, we separate $\hat{H}_I^{(n)}$ in terms of resonant (with subscript $0$), fast ($>$), and slow ($<$) Fourier modes set by a high-frequency cutoff $\Lambda_c$ satisfying $\lambda \le \Lambda_c \ll 1$.
Assuming $\hat{U}_\mathrm{fast}(s;\lambda) = \exp(-i \hat{\Phi}(s;\lambda))$,  we have the second-order QAT results
\begin{empheq}[box=\tightfbox]{equation}
    \begin{aligned}
        \hat{H}_\mathrm{eff}^{[2]}(s;\lambda) ={}& \sum_{n =1}^{2} \lambda^n \left(\hat{H}_{I,0}^{(n)} + \hat{H}^{(n)}_{I,<}(s) \right) \\
            - \lambda^2 \, \sum_{\mathclap{\substack{\Lambda_{\omega_k}> \Lambda_c \\ |\Lambda_{\omega_k} - \Lambda_{\omega_j}| \le \Lambda_c} }} & \, \Bigl( \frac{e^{-is(\Lambda_{\omega_k} - \Lambda_{\omega_j})}}{2\Lambda_{\omega_k}} [\hat{h}_{I,k},\hat{h}_{I,j}^\dagger] - h.c. \Bigr)
    \end{aligned}
\end{empheq}
and
\begin{empheq}[box=\tightfbox]{equation}
\begin{aligned}
    \hat{\Phi}^{[1]}(s;\lambda) ={}& \lambda \int^s ds' \hat{H}_{I,>}^{(1)}(s) \\
    ={}& \lambda \sum_{\mathclap{\Lambda_{\omega_k}> \Lambda_c}} \, \Bigl( \frac{e^{-is\Lambda_{\omega_k}}}{-i\Lambda_{\omega_k}} \hat{h}_{I,k} + h.c. \Bigr) 
\end{aligned}
\end{empheq}
such that $\hat{U}_I^{[2]}(s;\lambda) = \exp(-i\hat{\Phi}^{[1]}(s;\lambda)) \hat{U}_\mathrm{eff}^{[2]}(s;\lambda)$ approximates the exact solution with bounded error $\mathcal{O}(\lambda^2)$ at least over a time $s-s_0 \sim \mathcal{O}(1/\lambda)$.

To arrive at this result, this paper is organized as follows.
Section~\eqref{sec:qat} develops the theoretical QAT framework as applied to far-detuned systems, focusing on the integration of the unitarity-preserving Magnus expansion and the method of averaging on Hilbert spaces. 
The QAT framework yields an algorithmic procedure for generating $\hat{\Phi}^{(n)}(s)$ and $\hat{H}^{(n)}_\mathrm{eff}(s)$ to arbitrary $n$th order.
A discussion on error bounds and comparison to related Floquet-based methods is provided in Sec.~\eqref{sec:qat:validity}.
Section~\eqref{sec:qat_multiplescales} extends QAT to multi-timescale systems using two separate approaches, which are then shown to be equivalent: a two-timescale derivative expansion \cite{frascaUnifiedTheoryQuantum1992,frascaTheoryQuantumResonance1998,nayfehMethodMultipleScales2000} in Sec.~\eqref{sec:qat_multiplescales:twotime} and a Partitioned Expansion by Timescale Separation (PETS) approach in Sec.~\eqref{sec:qat_multiplescales:PETS}.
The latter yields the second-order results presented above and is applied to study entangling gate performance in the companion paper \cite{barajasMultiTimescaleCoherentControl2025}.
Finally, Section~\eqref{sec:conclusion} concludes with a summary of the key findings and prospects for future work.

\section{Quantum Averaging for Far-Detuned Interactions}\label{sec:qat}

In far-detuned systems ($\vec{\Lambda} \gg \lambda$), higher-order interactions often produce near-resonant beat notes, which terminate perturbative expansions through small denominator terms.
Valid all-order expansions exist only for time-periodic systems $\hat{H}_I(s) = \hat{H}_I(s+T)$ with periods $T < 2\pi/\lambda$ and a single drive frequency $\Lambda_\omega$ or harmonics $n\Lambda_\omega \in \vec{\Lambda} \; \forall \; n \in \mathbb{N}$ \cite{casasFloquetTheoryExponential2001,eckardtHighfrequencyApproximationPeriodically2015}.
Nonetheless, this section models far-detuned interactions under conditions where the approximation holds.
Extensions to near-resonant interactions using multi-timescale techniques are discussed in Sec.~\eqref{sec:qat_multiplescales}.

The standard MA formalism expands the fast propagator using a Dyson series (see \cite{schererNewPerturbationAlgorithms1997} and \cite{barajasMultiTimescaleCoherentControl2025}, appendix D).
The algebraic Dyson expansion, however, has numerous drawbacks for describing quantum dynamics (see \cite{blanesMagnusExpansionIts2009} and references therein); chiefly, the transformation is only unitary in the infinite resummation of the Dyson series.
Following unitary averaging frameworks \cite{casasFloquetTheoryExponential2001,casasTimedependentPerturbationTheory2015,eckardtHighfrequencyApproximationPeriodically2015}, we adopt a Lie theory of MA 
\cite{horiTheoryGeneralPerturbation1966,mersmanNewAlgorithmLie1970,schererNewPerturbationAlgorithms1997} with the symmetry-preserving exponential Lie transformation
\begin{equation}\label{eq:qat:QAT_exponential_transformation}
    \hat{U}_\mathrm{fast}(s;\lambda) = e^{-i \hat{\Phi}(s;\lambda)}, \quad \hat{\Phi}(s;\lambda) = \sum_{k=1}^{\infty} \lambda^k \hat{\Phi}^{(k)}(s)
\end{equation}
depending on a (Hermitian) dynamical phase operator $\hat{\Phi}(s;\lambda)$, which preserves unitarity upon truncation and has the simple inverse property $(e^{-i \hat{\Phi}})^{-1} = e^{i \hat{\Phi}}$ \footnote{The dynamical phase operator is also referred to as a Magnus exponent or micromotion operator in the literature.}.
Setting $\hat{\Phi}^{(0)} = 0$ ensures $\lim_{\lambda \to 0} \hat{U}_{I}(s;\lambda) = \idty$ in the absence of the perturbation.

While inserting \eqref{eq:qat:QAT_exponential_transformation} into \eqref{eq:qat:interaction_eq} yields a complex non-linear differential equation (see appendix~\eqref{appendix:Magnus_type_expansions}), the group-preserving Lie approach balances complexity with computational efficiency and enhanced predictability via the Magnus expansion \cite{magnusExponentialSolutionDifferential1954, blanesMagnusExpansionIts2009}.
To systematically solve the effective interaction picture, we use an iterative Picard scheme, which generates the QAT homological equation
\begin{equation}\label{eq:qat:QAT_homological}
	\frac{d}{ds}\hat{\Phi}^{(n)}(s) = \hat{\mathcal{H}}_{\Phi}^{(n)}(s) - \hat{H}_{I,\mathrm{eff}}^{(n)}, \quad n \ge 1
\end{equation}
where the auxiliary Hamiltonian operator is defined as
\begin{equation}\label{eq:qat:QAT_aux_generator}
	\hat{\mathcal{H}}_{\Phi}^{(n)} = \hat{H}_I^{(n)} + \sum_{k=1}^{n-1} \frac{B_k}{k!} \, \left((-)^k \,  \hat{S}_{k}^{(n)} - \hat{T}_{k}^{(n)} \right)
\end{equation}
where $B_k$ are the Bernoulli numbers.
The operators $\hat{S}_{k}$ and $\hat{T}_{k}$ are generated by the recurrence relations
\begin{subequations}\label{eq:qat:QAT_aux_recurrence}
    \begin{align}
	    \hat{S}_{0}^{(n)} &= \hat{H}_{I}^{(n)}, \quad
     \hat{T}_{0}^{(n)} = \hat{H}_{I,\mathrm{eff}}^{(n)} \\
	    \hat{A}_{k}^{(n)} &= \sum_{m=1}^{n-k} \left[ i \hat{\Phi}^{(m)},\hat{A}_{k-1}^{(n-m)} \right], \quad 1 \le k \le n-1
    \end{align}
\end{subequations}
with $\hat{A}$ replaced by $\hat{S}$ or $\hat{T}$ with explicit time-dependence omitted and the adjoint action $\mathrm{ad}_{\hat{X}}(\hat{Y}) = [\hat{X},\hat{Y}]$ and $\mathrm{ad}_{\hat{X}}^{(k)}(\hat{Y}) = [\hat{X},\mathrm{ad}_{\hat{X}}^{(k-1)}(\hat{Y})]$ for integer $k \ge 2$.

The auxiliary operator, $\hat{\mathcal{H}}_{\Phi}^{(n)}(s)$, is introduced to subsume all combinations of $\hat{H}_I^{(k)}$, $\hat{\Phi}^{(k)}$, and $\hat{H}^{(k)}_{I,\mathrm{eff}}$ for $1 \le k \le n-1$ in a convenient form with the first three terms provided in Table~\eqref{table:qat:auxiliary_hamiltonian}.
The recurrence relation provides a straightforward implementation in a symbolic solver package that reduces computation time by recycling previously calculated terms.
% == Table I == %
\begin{table}[t]
\caption{QAT Auxiliary Hamiltonian at Different Orders}\label{table:qat:auxiliary_hamiltonian}

% \resizebox{\columnwidth}{!}{%
% \setlength{\tabcolsep}{12pt} % Default value: 6pt
% \renewcommand{\arraystretch}{2.25} % Default value: 1
\centering
\begin{ruledtabular}
\begin{tabular}{rr@{}l}
% \hline\hline
% \toprule\toprule
    % \multicolumn{3}{c@{}}{\text{Randbetingelse}} \\
% \cmidrule(c){1-3}
% \colrule
    Order & \multicolumn{2}{l@{}}{\text{Definition}} \\
    \hline
    \addlinespace
    $n=1,$ & $\hat{\mathcal{H}}_{\Phi}^{(1)}(s) =$  & $ \hat{H}_I^{(1)}(s)$\\
    \addlinespace\addlinespace
    $n=2,$ & $\hat{\mathcal{H}}_{\Phi}^{(2)}(s) =$   &  $ \hat{H}_I ^{(2)}(s) + \frac{1}{2} [i \hat{\Phi}^{(1)}(s), \hat{H}_I ^{(1)}(s) + \hat{H}{\vphantom{H}}^{(1)}_{I,\mathrm{eff}}] $\\
    \addlinespace\addlinespace
    % $n=3,$ & $\hat{\mathcal{H}}_{\Phi}^{(3)}(s)=$   & \parbox[][][l]{0cm}{ \begin{equation*}\begin{aligned}
    %     &\hat{H}_I^{(3)} + \frac{1}{2} \left( [i \hat{\Phi} ^{(2)}, \hat{H}_I^{(1)} + \hat{\mathcal{H}}{\vphantom{H}}^{(1)}_{I,\mathrm{eff}}] \right. \\
    %     &\quad \left. + [i \hat{\Phi} ^{(1)}, \hat{H}_I^{(2)} +\hat{\mathcal{H}}{\vphantom{H}}^{(2)}_{I,\mathrm{eff}}] \right) \\
    %     &\quad - \frac{1}{12} \left( [i \hat{\Phi} ^{(1)},[i \hat{\Phi}^{(1)},\hat{\mathcal{H}}{\vphantom{H}}^{(1)}_{I,\mathrm{eff}}-\hat{H}_I^{(1)}]] \right )
    % \end{aligned}\end{equation*}}
    $n=3,$ & $\hat{\mathcal{H}}_{\Phi}^{(3)}(s)=$   & $\hat{H}_I^{(3)} + \frac{1}{2} \left( [i \hat{\Phi} ^{(2)}, \hat{H}_I^{(1)} + \hat{H}{\vphantom{H}}^{(1)}_{I,\mathrm{eff}}] \right.$ \\
    \addlinespace
        &   & $\quad \left. + [i \hat{\Phi} ^{(1)}, \hat{H}_I^{(2)} +\hat{H}{\vphantom{H}}^{(2)}_{I,\mathrm{eff}}] \right)$ \\
    \addlinespace
        &   & $\quad - \frac{1}{12} \left( [i \hat{\Phi} ^{(1)},[i \hat{\Phi}^{(1)},\hat{H}{\vphantom{H}}^{(1)}_{I,\mathrm{eff}}-\hat{H}_I^{(1)}]] \right )$
% \bottomrule
\end{tabular}
\end{ruledtabular}
% }
\end{table} 

The effective Hamiltonian, unknown \textit{a priori}, remains flexible but must regularize the homological equation to avoid secular growth.
From the method of averaging, we introduce the simple time-averaging procedure
\begin{equation}\label{eq:qat:QAT_time_average}
	\timeavg{\hat{A}^{(n)}}{s} = \lim_{T \to \infty} \frac{1}{T} \int_{{s_0}}^{{T+s_0}} {\hat{A}^{(n)}(s)} \: d{s} {},
\end{equation}
which is valid only if the limit uniformly exist independent of the choice of $s_0$ and additional parameters. 
Then eq.~\eqref{eq:qat:QAT_homological} is regularized by systematically requiring
\begin{equation}\label{eq:qat:QAT_phase_cond}
	\timeavg{\hat{\Phi}^{(n)}(s)}{s} = 0,
\end{equation}
which is satisfied by the \textit{time-independent} effective Hamiltonian
\begin{equation}\label{eq:qat:QAT_effHam_cond}
	\hat{H}_{I,\mathrm{eff}}^{(n)} = \timeavg{\hat{\mathcal{H}}_{\Phi}^{(n)}(s)}{s}
\end{equation}
that removes the appearance of false secular terms from resonant interactions and ensuring the dynamical phase remains bounded \footnote{Recall from \eqref{eq:qat:PT_first_order} that the appearance of secular terms with unbounded growth as $s \rightarrow \infty$ lead to breakdown of the asymptotic analysis for long-time dynamics.}.
Less obvious is that the regularization condition requires that $\timeavg{\hat{U}_I(s;\lambda)}{s} = \hat{U}_\mathrm{eff}(s;\lambda)$.
Finally, solving the homological equation yields
\begin{equation}\label{eq:qat:QAT_homological_integral}
	\hat{\Phi}^{(n)}(s) = \int^s ds' \left( \hat{\mathcal{H}}_{\Phi}^{(n)}(s') - \hat{H}_{I,\mathrm{eff}}^{(n)} \right)
\end{equation}
where the indefinite integral enforces a uniqueness rule for the iterative Picard sequence and ensures the expansion is manifestly gauge invariant. 
Enforcing a uniqueness rule amounts to appropriately choosing the integration constant in a manner that can be systematically applied to all orders.
We use the Van Vleck gauge defined as the integration constant set to zero, which is to be assumed when not explicitly stated.
The choice may limit the possible transformations guided by $\hat{\Phi}$, but ensures a unique asymptotic expansion (up to a gauge transformation).
For example, a definite integral with a lower integration bound $s'=s_0$ corresponds to the Magnus gauge, a non-gauge invariant uniqueness rule with integration constant $-\hat{\Phi}^{(n)}(s_0)$ that instead satisfies $\partial_s \timeavg{\hat{\Phi}^{(n)}(s)}{s} = 0$.

The exponential Lie approach simplifies quantum averaging by framing slow effective dynamics as riding in phase with fast-varying effects, akin to surfing atop ocean waves.
From the point of view of the renormalization group (RG) method \cite{chenRenormalizationGroupSingular1996,frascaRenormalizationGroupMethods1997,frascaTheoryQuantumResonance1998}, $\hat{H}_{I,\mathrm{eff}}^{(n)}$ is regarded as a regularization parameter that allows us to renormalize the Magnus expansion.
Moreover, the effective interaction picture equation that describes the slowly varying dynamics generated by $\hat{H}_{I,\mathrm{eff}}(\lambda)$ is simply the RG equation. 
Therefore, QAT returns a \textit{renormalized} Magnus expansion for the interaction propagator, factorized into the non-secular unitary fast propagator characterizing only fast-varying dynamics modulating the effective propagator solution to the RG equation \cite{zianeCertainRenormalizationGroup2000}.

To implement QAT, we approximate \eqref{eq:qat:interaction_propagator_factorized} using algorithm~\eqref{fig:QAT_algorithm}, which systematically calculates the effective Hamiltonian and dynamical phase operator to the desired order of precision.
\begin{algorithm}[tb]\label{fig:QAT_algorithm}
\caption{QAT Procedure}
% \normalsize
Let $n = 1$\;
\While{$n\leq N$}
{
    \vspace{2pt}
    Compute $\hat{S}_k^{(n)}$ and $\hat{T}_k^{(n)}$ for $1 \le k \le n-1$\;
    \vspace{5pt}
    Compute $\hat{\mathcal{H}}^{(n)}_{\Phi}(s)$ from $\{\hat{S}_k^{(n)},\hat{T}_k^{(n)}\}_{k=1}^{n-1}$\;
    \vspace{5pt}
    Apply the regularization condition $\timeavg{\hat{\mathcal{H}}^{(n)}_{\Phi}(s)}{s} = \hat{H}_{I,\mathrm{eff}}^{(n)}$\;
    \vspace{5pt}
    \If{$n\leq N-1$}
    {
        \vspace{2pt}
        Integrate $\partial_s\hat{\Phi}^{(n)}(s) = \hat{\mathcal{H}}^{(n)}_{\Phi}(s) - \hat{H}_{I,\mathrm{eff}}^{(n)}$\;
        \vspace{2pt}
    }
}
\end{algorithm}
Arriving at the desired order, one has the following asymptotic approximations:
\begin{subequations}\label{eq:qat:QAT_asymptotic_approx}
    \begin{align}
        \hat{H}^{[N]}_{I,\mathrm{eff}}(\lambda) ={}& \sum_{n=1}^{N} \lambda^n \, \hat{H}^{(n)}_{I,\mathrm{eff}} \\
        \hat{\Phi}^{[N-1]}(s;\lambda) ={}& \sum_{n=1}^{N-1} \lambda^n \, \hat{\Phi}^{(n)}(s)
    \end{align}
\end{subequations}
where square brackets indicate the truncation order (note the $N$th dynamical phase contribution is not strictly necessary).
With \eqref{eq:qat:QAT_eff_system} solved for $\hat{H}_{I,\mathrm{eff}} \simeq \hat{H}^{[N]}_{I,\mathrm{eff}}$, then $\hat{U}_I(s;\lambda)$ is approximated by the truncated QAT solution
\begin{equation}\label{eq:qat:QAT_interaction_approx}
	\hat{U}^{[N]}_I(s;\lambda) = \hat{U}_\mathrm{fast}^{[N-1]}(s;\lambda) \hat{U}^{[N]}_\mathrm{eff}(\tau;\lambda)|_{\tau = \lambda s}
\end{equation}
where 
\begin{subequations}
\begin{align}
    \hat{U}_\mathrm{fast}^{[N-1]}(s;\lambda) ={}& \exp(-i \hat{\Phi}^{[N-1]}(s;\lambda)) \\
    \hat{U}_\mathrm{eff}^{[N]}(\tau;\lambda) ={}& \exp(-i \sum_{n=0}^{N-1} \lambda^{n}\hat{H}_{I,\mathrm{eff}}^{(n+1)} \, \tau)
\end{align}
\end{subequations}
where $\tau = \lambda t$ is the previously introduced slow timescale and the method is complete.
Note that \eqref{eq:qat:QAT_eff_system} can also be approximated with time-independent perturbation theory up to $\mathcal{O}(\lambda^N)$ and is easily solved numerically.
We remark that the far-detuned QAT results can be equivalently obtained with prior quantum averaging methods \cite{schererNewPerturbationAlgorithms1997,casasFloquetTheoryExponential2001,casasTimedependentPerturbationTheory2015,arnalExponentialPerturbativeExpansions2020}.
However, the explicit connection to the method of averaging allows us to generalize the result to almost-periodic systems, distinguish previous Magnus-type methods based on the choice of uniqueness rule, and facilitate a multi-timescale treatment in Sec.~\eqref{sec:qat_multiplescales}.

\subsection{On Validity, Error Bounds, and Related Time-Averaged High-Frequency Expansions}\label{sec:qat:validity}

Despite extensive research on quantum averaging and effective Hamiltonian theories, a review of validity conditions, error bounds, and relation between Magnus-based Floquet methods are believed to be a valuable contribution.
Recent findings \cite{maggiaHigherorderAveragingTimeperiodic2020,bauerTimeorderingGeneralizedMagnus2013} confirm the self-consistency of algebraic and Lie-theoretic transformation approaches allowing us to recite the following classical averaging results.
A critical validity condition is the uniform definition of time averaging yielding \cref{eq:qat:QAT_phase_cond,eq:qat:QAT_effHam_cond} and the absence of small denominator terms \cite{eckhausNewApproachAsymptotic1975}.
For arbitrary bounded perturbations there is no guarantee that the homological equation remains uniformly bounded to all orders.
The first-order approximation provides weak error bounds for general bounded perturbations with a well-defined time average: 
the approximate solution, $\hat{U}_{I,\mathrm{eff}}^{[1]}(s;\lambda)$, to the truncated time-averaged equation
\begin{equation}
	\frac{d}{ds}\hat{U}_{I,\mathrm{eff}}(s;\lambda) = \lambda \, \hat{H}_{I,\mathrm{eff}}^{(1)} \, \hat{U}_{I,\mathrm{eff}}(s;\lambda) + \mathcal{O}(\lambda^2)
\end{equation}
with initial condition $\hat{U}_{I,\mathrm{eff}}(0) = \hat{U}_I(0)$ leads to the well-known first-order averaging result
\begin{equation}
	\lVert \hat{U}_I(s) - \hat{U}_{I,\mathrm{eff}}^{[1]}(s) \rVert =
	\begin{cases}
		\mathcal{O}(\lambda), & \text{if almost-periodic} \\
		\mathit{o}(1), & \text{otherwise}
	\end{cases}
\end{equation}
applicable for times $s-s_0 \sim \mathcal{O}(1/\lambda)$ \cite{eckhausNewApproachAsymptotic1975,buitelaarMethodAveragingBanach1993,sandersAveragingMethodsNonlinear2007}.
However, the QAT approximation is guaranteed to be valid to all orders for periodic and almost-periodic Fourier systems \cite{finkAlmostPeriodicDifferential1974,levitanAlmostPeriodicFunctions1983} on finite and infinite dimensional Hilbert spaces \cite{buitelaarMethodAveragingBanach1993}.
For these systems, we have the following higher-order error bounds \cite{perkoHigherOrderAveraging1969}: for each consecutive $\hat{H}_{I,\mathrm{eff}}^{(j)} = 0$ for $0 \le k < N$, the approximate solution $\hat{U}_I^{[N]}$ in eq.~\eqref{eq:qat:QAT_interaction_approx} is bounded near the exact solution $\hat{U}_I$ by
    \begin{equation}\label{eq:qat:error_nth_approx}
			\lVert \hat{U}_I(s;\lambda) - \hat{U}_{I}^{[N]}(s;\lambda) \rVert = \mathcal{O}(\lambda^{N-k})
    \end{equation}
for a time $T_\lambda = s-s_0 \sim \mathcal{O}(1/\lambda^{k+1})$ and $\lambda$ sufficiently small (hence the omission of $\hat{\Phi}^{(N)}$ in \eqref{eq:qat:QAT_asymptotic_approx}). 
Therefore, $\hat{U}_{I,\mathrm{eff}}$ must start and remain near the true solution trajectory over the finite timespan $T_\lambda$, which means $\hat{H}_{I,\mathrm{eff}}$ is guaranteed to describe the slowly-varying dynamics during that finite time \cite{perkoHigherOrderAveraging1969}.
These bounds align with the sufficient condition for guaranteed convergence of Magnus-type expansions for bounded operators on Hilbert spaces,
\begin{equation}
    \int_{s_0}^s \lVert \hat{H}_I(s') \rVert_2 \, ds' < \pi
\end{equation}
where $\lVert \blank \rVert_2 = \max_{|\braket{\psi | \psi}|=1} \lVert \blank \ket{\psi} \rVert_2 $ is the 2-norm \cite{casasSufficientConditionsConvergence2007}.

In $T$-periodic systems, QAT aligns with Floquet’s theorem, producing solutions in the form
\begin{equation}\label{eq:qat:Floquet_unitary}
	\hat{U}_I(s) \doteq  \hat{U}_{\mathrm{fast}}(s) \, e^{-i \hat{H}_\mathrm{eff} s}, \quad \hat{U}_{\mathrm{fast}}(s) = \hat{U}_{\mathrm{fast}}(s+T) %\exp \left( \hat{\mathfrak{F}}(s) \right)
\end{equation}
where $\hat{H}_\mathrm{eff}$ is the time-independent and $s_0$-invariant Floquet Hamiltonian and $\hat{U}_{\mathrm{fast}}(s)$ is the $T$-periodic fast propagator (also known as the ``micromotion'' operator) \cite{casasFloquetTheoryExponential2001,bukovUniversalHighFrequencyBehavior2015}.
Periodic QAT results, manifestly gauge invariant in the Van Vleck gauge \cite{eckardtHighfrequencyApproximationPeriodically2015,goldmanPeriodicallyDrivenQuantum2014}, contrast with the Floquet-Magnus expansion, where gauge choices introduce $s_0$-dependence \cite{casasFloquetTheoryExponential2001, kuwaharaFloquetMagnusTheory2016}.
The latter yields the time-centered propagator $\hat{U}_I(s,s_0) = \hat{U}_{\mathrm{fast}}(s,s_0) \, e^{-i \hat{H}^{\mathrm{FM}}_\mathrm{eff}[s_0] (s-s_0)}$ for an $s_0$-dependent Floquet-Magnus effective Hamiltonian $\hat{H}^{\mathrm{FM}}_\mathrm{eff}[s_0]$.
%, as opposed to the $s_0$-invariant propagator in \eqref{eq:qat:Floquet_unitary}.
The difference in these two expansions can be attributed to the choice of the uniqueness rule in \eqref{eq:qat:QAT_homological_integral} with the latter including the integration constant dependent on $s_0$.
The gauge relation between the two expansion can be expressed as
\begin{equation}
    \hat{H}_{\mathrm{eff}}^{\mathrm{FM}}[s_0] = \hat{U}_\mathrm{fast}(s_0) \hat{H}_\mathrm{eff} \hat{U}_\mathrm{fast}^\dagger(s_0)
\end{equation}
and identifying $\hat{U}_{\mathrm{fast}}(s,s_0) = \hat{U}_{\mathrm{fast}}(s)\hat{U}^\dagger_{\mathrm{fast}}(s_0)$.
In particular, Ref.~\cite{eckardtHighfrequencyApproximationPeriodically2015} shows that the spurious $s_0$-dependence leads to inaccurate predictions of the approximate quasienergy spectrum of the Floquet Hamitlonian emphasizing the utility of the $s_0$-invariant Van Vleck gauge.
As previously noted, the method applies to almost-periodic operators until small denominator terms appear, aligning with generalized Floquet-Magnus theories for high-frequency perturbations \cite{finkAlmostPeriodicDifferential1974,verdenyQuasiPeriodicallyDrivenQuantum2016,arnalExponentialPerturbativeExpansions2020}. 
Finally, the QAT framework extends to bounded, quasi-periodic pulse sequences that admit a convergent infinite-dimensional Fourier decomposition
%the method typically holds for quasi-bounded perturbations (at least bounded for some finite time) expanded in a generalized Fourier series 
$\hat{H}_I(s) = \sum_{k=-\infty}^{\infty} e^{i \Lambda_k s} \, \hat{H}_I^{(k)}$ with dimensionless frequency $\Lambda_k > 0$ independent of $\lambda$, $ \hat{H}_I^{(-k)} = \hat{H}_I^{(k) \, \dagger}$ and $\Lambda_{-k} = - \Lambda_k$, and mean-free $\hat{H}_I^{(0)}=\timeavg{\hat{H}_I}{s}$ \cite{finkAlmostPeriodicDifferential1974}.
In practice, even unbounded ramps (e.g., in Landau–Zener-type protocols) can be approximated as effectively bounded over finite durations, allowing their decomposition into a quasi-periodic series for the relevant time window.
As long as the driving field admits a well-defined spectral separation between fast and resonant modes, far-detuned QAT remains applicable.
Hence, far-detuned QAT is a generalized Floquet-Magnus theory for high-frequency, almost-periodic perturbations.

Finally, while QAT may resemble regular perturbation theory, it is fundamentally distinct.
In perturbation theory, the interaction picture propagator $\hat{U}_I(s;\lambda)$ is asymptotically expanded, directly solving for dynamics through typically secular terms valid only for short times.
In contrast, QAT constructs a gauge transformation into the effective interaction picture, where $\hat{H}_{I,\mathrm{eff}}$ mitigates secular growth and approximates the system's qualitative long-time behavior.
The truncated form of eq.~\eqref{eq:qat:QAT_interaction_approx} captures the dynamics of $\hat{H}_I$ without relying on regular perturbation theory.
If $\hat{H}_{I,\mathrm{eff}}^{[N]}$ is not explicitly solvable, perturbation theory may be employed as a secondary step (see Appendix~\eqref{appendix:effective_equation_PT}).
We now illustrate the QAT expansion with the semi-classical Rabi problem, a paradigmatic example of a periodically driven quantum system.

\subsubsection*{Example 1: The Far-Detuned Rabi Problem}\label{sec:qat:ex_Rabi}
Consider a two-level system (2LS) that weakly interacts with an applied AC electromagnetic field, periodically driven at frequency $\omega$ \cite{sakuraiModernQuantumMechanics2020}.
When resonantly driven, the AC field slowly induces population inversion between the ground and the excited state, which can be mapped onto a basic logical operation for quantum computation \cite{nielsenQuantumComputationQuantum2010}.
In particular, there is an exact solution for the complex-valued perturbation $\hat{V}(t)/\hbar = \frac{\Omega}{2} \left( e^{-i \omega t} \, \hat{\sigma}_+ + h.c. \right)$ suitable for comparing with the performance of the QAT results. 
The total Hamiltonian in the Schr\"odinger picture is given by
\begin{equation}\label{eq:ex:Rabi_total_Hamiltonian}
	\hat{H}_\mathrm{S}(t)/\hbar
 = \frac{\omega_{eg}}{2} \, \hat{\sigma}_z + \frac{\Omega}{2} \left( e^{-i \omega t} \, \hat{\sigma}_+ + h.c. \right),
\end{equation}
where $\hbar \omega_{eg}$ is the energy gap between the ground ($\ket{1}$) and the excited ($\ket{0}$) state with the perturbation amplitude characterized by the resonant Rabi frequency $\Omega$.
The states are directly coupled by the perturbation with the raising operator $\hat{\sigma}_{+} = \ket{0}\bra{1}$ and lowering operator $\hat{\sigma}_{-} = \ket{1}\bra{0}$.
The total Hamiltonian admits the closed-form solution (for $t_0 = 0$)
\begin{equation}\label{eq:ex:Rabi_full_solution}
	\hat{U}_\mathrm{total}(t) = e^{-i \hat{H}_0 t /\hbar} \, e^{-i \Delta \hat{\sigma}_z t/2} \, e^{-i \, \left(-\Delta \, \hat{\sigma}_z + \Omega \, \hat{\sigma}_x \right) t/2 }  
\end{equation}
where the detuning parameter $\Delta = \omega - \omega_{eg}$ determines the resonance condition.

For present purposes we study the system weakly driven far from resonance, setting the stage for the broader discussion of the role of far-detuned and near-resonant interactions within multi-timescale dynamics in \cref{sec:qat_multiplescales}.
The perturbative regime is defined by the small parameter
\begin{equation}
   \lambda = \frac{\Omega/2}{\omega_{eg}} \ll 1
\end{equation}
following the usual quantum optics convention.
Before proceeding, we place \eqref{eq:ex:Rabi_total_Hamiltonian} in the dimensionless form of \eqref{eq:qat:SchroHamiltonian} with the scaled time $s = \omega_{eg} t$.
Upon rescaling, we have $\hat{H}_0 = \hat{\sigma}_z / 2$ and $\hat{V}(s;\lambda) = \lambda \left( e^{-i \Lambda_\omega s} \, \hat{\sigma}_+ + h.c. \right)$ where $\Lambda_\omega = \omega/\omega_{eg}$.

\subparagraph{Interaction Picture.} 
Expressing the perturbation as $\hat{V}(s;\lambda) = \lambda \, \hat{V}^{(1)}(s)$ for $\hat{V}^{(1)}(s) = \left( e^{-i \Lambda_\omega s} \, \hat{\sigma}_+ + h.c. \right)$, in the interaction picture with respect to $\hat{H}_0$ we have $\hat{H}_I(s;\lambda) = \lambda \, \hat{H}_I^{(1)}(s)$ where
\begin{equation}\label{eq:ex:Rabi_interaction_hamiltonian}
\begin{aligned}
	\hat{H}_I^{(1)}(s) &= e^{i \hat{H}_0 s} \, \hat{V}^{(1)}(s;\lambda) \, e^{-i \hat{H}_0 s} \\
 &= e^{-i \Lambda_\Delta s} \, \hat\sigma_+ + h.c.
\end{aligned}
\end{equation}
with the (dimensionless) detuning $\Lambda_\Delta = \Delta/\omega_{eg}$ and $|\Lambda_\Delta|\gg \lambda$.
The interaction Hamiltonian is already in the Fourier form of \eqref{eq:qat:Fourier_basis}, thereby guaranteed to be compatible with the QAT formalism.

\subparagraph{QAT Homological Equation.}
We proceed with the QAT algorithm, repeated here for convenience: calculate $\hat{\mathcal{H}}^{(n)}_{\Phi}(s)$ and $\hat{H}_{I,\mathrm{eff}}^{(n)} = \timeavg{\hat{\mathcal{H}}{\vphantom{H}}^{(n)}_{\Phi}(s)}{s}$, then integrate \eqref{eq:qat:QAT_homological_integral} for $\hat{\Phi}^{(n)}(s)$. 
From Table~\ref{table:qat:auxiliary_hamiltonian} and eq.~\eqref{eq:qat:QAT_time_average} for $n=1$ we have
\begin{equation}
    \begin{aligned}
            \hat{\mathcal{H}}_{\hat{\Phi}}^{(1)}(s) ={}& \hat{H}_I^{(1)}(s) \\
	    \timeavg{\hat{\mathcal{H}}_{\Phi}^{(1)}(s)}{s} ={}& \lim_{T \to \infty} \frac{1}{T}\int_{{0}}^{{T}} {\hat{H}_I^{(1)}(s)} \: d{s} = 0
     \end{aligned}
\end{equation}
leading to the first-order contribution
\begin{subequations}
	\begin{align}
        \hat{H}_{I,\mathrm{eff}}^{(1)} ={}& \timeavg{\hat{\mathcal{H}}_{\Phi}^{(1)}(s)}{s} = 0 \label{eq:ex:Rabi_QAT_eff_o1} \\
        \begin{split} \label{eq:ex:Rabi_QAT_dynphase_o1}
		\hat{\Phi}^{(1)}(s) ={}&  \int^s d{s'} \: {\left(\hat{\mathcal{H}}_{\hat{\Phi}}^{(1)}(s') - \hat{H}_{I,\mathrm{eff}}^{(1)}\right)} \\
        ={}& \frac{1}{-i \Lambda_\Delta} \exp(-i \Lambda_\Delta s) \, \hat{\sigma}_+ + h.c.
        \end{split}
	\end{align}
\end{subequations}
Repeating the process at next order returns
\begin{subequations}
	\begin{align}
        \hat{H}_{I,\mathrm{eff}}^{(2)} ={}& \timeavg{\hat{\mathcal{H}}_{\Phi}^{(2)}(s)}{s} = -\frac{1}{\Lambda_\Delta} \, \hat{\sigma}_z \\
        \begin{split}
	\hat{\Phi}^{(2)}(s) ={}& \int^s ds' \: {\left(\hat{\mathcal{H}}_{\hat{\Phi}}^{(2)}(s') - \hat{H}_{I,\mathrm{eff}}^{(2)}\right)} = 0,
        \end{split}
	\end{align}
\end{subequations}
which can easily be verified.
Since $\hat{H}_{I,\mathrm{eff}}^{(2)} \propto \hat{\sigma}_z$ is diagonal, the effective interaction due to the far-detuned driving generates a shift in resonance frequency known as the Bloch-Siegert shift in NMR spectroscopy and a light shift in quantum optics.

%% == PLOT == %%
\begin{figure}[!t]
    \centering
    % \includegraphics[width=8.6cm]{plots/Raman_plot.pdf}
    %\resizebox{1.01\columnwidth}{!}{
    \includestandalone[width=\columnwidth]{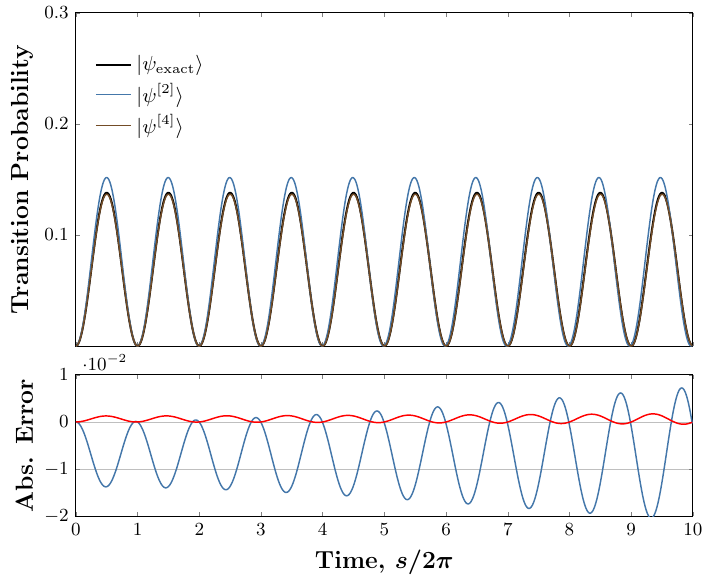}
    \caption{
    \textbf{Far-Detuned Rabi Oscillations.}  
        The system is initialized in the ground state $\ket{g}$ and driven with a weak-field amplitude $\lambda = 0.5 \times 10^{-5}$.  
        For a valid high-frequency QAT expansion, the detuning must satisfy $|\Lambda_\Delta| > \lambda$, with improved convergence as $\lambda/|\Lambda_\Delta| \ll 1$.  
        To illustrate this, we choose a moderately large ratio $\lambda/|\Lambda_\Delta| = 0.4$.  
        Second-order (blue) and fourth-order (red) QAT results are shown to match the envelope of the exact solution, with the fourth-order expansion exhibiting strong agreement over multiple cycles.
    }
    \label{fig:RabiFarDetuned}
\end{figure}

\subparagraph{The Effective Interaction Picture.}
Truncating the iterative process at second-order yields
\begin{equation}
	\hat{H}_{I,\mathrm{eff}}^{[2]}(\lambda) = \sum_{j=1}^{2} \lambda ^{j} \, \hat{H}_{I,\mathrm{eff}}^{(j)} = \lambda^2 \left( -\frac{1}{\Lambda_\Delta} \, \hat{\sigma}_z \right)
\end{equation}
governing the effective interaction picture equation for $\hat{H}_{I,\mathrm{eff}} \simeq \hat{H}_{I,\mathrm{eff}}^{[2]}$ in \eqref{eq:qat:QAT_eff_system}.
The truncated effective propagator is trivially returned by
\begin{equation}\label{eq:ex:Rabi_effective_propagator}
\begin{aligned}
	\hat{U}_\mathrm{eff}^{[2]}(s;\lambda) ={}& \exp \left(-i \, \hat{H}_{I,\mathrm{eff}}^{[2]}(\lambda) \, s \right) \\
        ={}& \cos(\lambda^2 s/\Lambda_\Delta) \idty + i \sin(\lambda^2 s/\Lambda_\Delta) \hat{\sigma}_z.
\end{aligned}
\end{equation}

\subparagraph{Approximate QAT Dynamics.}
Finally, from eq.~\eqref{eq:qat:QAT_interaction_approx}, the interaction propagator is approximated by the truncated unitary QAT result
\begin{equation}\label{eq:ex:Rabi_interaction_propagator_approx}
\begin{aligned}
	\hat{U}_I^{[2]}(s) ={}& \hat{U}_{\mathrm{fast}}^{[1]}(s;\lambda) \hat{U}_{\mathrm{eff}}^{[2]}(s;\lambda) \\
	={}& \cos(\lambda / \Lambda_\Delta) e^{i \frac{\lambda^2}{\Lambda_\Delta } \hat{\sigma}_z s} \\
        &\quad + \sin(\lambda/\Lambda_\Delta)  (e^{- i \tilde{\Lambda}_\Delta s} \hat{\sigma}_{+} - h.c. )
\end{aligned}
\end{equation}
where $\tilde{\Lambda}_\Delta = \Lambda_\Delta + \frac{\lambda^2}{\Lambda_\Delta }$ and
\begin{equation}
\begin{aligned}
\hat{U}_\mathrm{fast}^{[1]}(s;\lambda) ={}& \exp(-i \hat{\Phi}^{[1]}(s;\lambda)) \\
={}& \cos(\lambda/\Lambda_\Delta) \idty \\
&\quad + \sin(\lambda /\Lambda_\Delta) ( e^{-i \Lambda_\Delta s} \hat{\sigma}_{+} - h.c.)
\end{aligned}
\end{equation}
with bounded error estimate $\mathcal{O}(\lambda)$ valid at least over a time $s\sim \mathcal{O}(1/\lambda^2)$ since $\hat{H}_{I,\mathrm{eff}}^{(1)} = 0$.
From these results, the approximate quantum state in the interaction picture, initialized in $\ket{\psi_0}$ at time $s=0$, is returned by
\begin{equation}
\begin{aligned}
    \ket{\psi_I^{[2]}(s,s_0=0)} ={}& \hat{U}_I^{[2]}(s) \hat{U}_I^{\dagger [2]}(s_0) \ket{\psi_0} \\
    \equiv{}& \hat{U}_I^{[2]}(s,s_0) \ket{\psi_0}
\end{aligned}
\end{equation}
where $\hat{U}^{[2]}_I(s_0,s_0) = \mathds{1}$ as expected.
While population is never transferred in the effective interaction picture when initialized in an eigenstate of $\hat{\sigma}_z$ (i.e. $\ket{\psi_0} = \ket{0}$ or $\ket{1}$), that is not the case when including the fast-varying effects carried within the dynamical phase.
The first-order dynamical phase contribution weakly generates an off-resonant transition with suppressed probability amplitude $\sin(\lambda/\Lambda_\Delta) \approx \lambda/\Lambda_\Delta$.
Figure~\eqref{fig:RabiFarDetuned} shows that by including the fast dynamics captured by the QAT framework we properly account for the non-negligible, off-resonant population transfer that appears in the numerical simulation.

% == Table II == %
\begin{table}[t]\label{table:FarDetuned_Rabi_Problem}
    \caption{Far-Detuned Rabi Problem: Effective Hamiltonian and Dynamical Phase Contributions at Different Orders}
    \centering
    \renewcommand{\arraystretch}{2}
    \begin{ruledtabular}
    \begin{tabular}{l l}
        $ \hat{H}_{I,\text{eff}}^{(1)} \,=\, 0 $ 
        & $ \hat{\Phi}^{(1)} \,=\, -\frac{1}{\Lambda_{\Delta}} \left( i e^{-i \Lambda_{\Delta} s} \hat{\sigma}_+ + h.c. \right) $ \\
        $ \hat{H}_{I,\text{eff}}^{(2)} \,=\, -\frac{1}{\Lambda_{\Delta}} \hat{\sigma}_z $ 
        & $ \hat{\Phi}^{(2)} \,=\, 0 $ \\
        $ \hat{H}_{I,\text{eff}}^{(3)} \,=\, 0 $ 
        & $ \hat{\Phi}^{(3)} \,=\, \frac{4}{3} \frac{1}{\Lambda_{\Delta}^3} \left( i e^{-i \Lambda_{\Delta} s} \hat{\sigma}_+ + h.c. \right) $ \\
        $ \hat{H}_{I,\text{eff}}^{(4)} \,=\, \frac{1}{\Lambda_{\Delta}^3} \hat{\sigma}_z $ 
        & $ \hat{\Phi}^{(4)} \,=\, 0 $ \\
        $ \hat{H}_{I,\text{eff}}^{(5)} \,=\, 0 $ 
        & $ \hat{\Phi}^{(5)} \,=\, -\frac{16}{5} \frac{1}{\Lambda_{\Delta}^5} \left( i e^{-i \Lambda_{\Delta} s} \hat{\sigma}_+ + h.c. \right) $ \\
        $ \hat{H}_{I,\text{eff}}^{(6)} \,=\, -\frac{2}{\Lambda_{\Delta}^5} \hat{\sigma}_z $ 
        & $ \hat{\Phi}^{(6)} \,=\, 0 $ \\
        $ \hat{H}_{I,\text{eff}}^{(7)} \,=\, 0 $ 
        & $ \hat{\Phi}^{(7)} \,=\, \frac{64}{7} \frac{1}{\Lambda_{\Delta}^7} \left( i e^{-i \Lambda_{\Delta} s} \hat{\sigma}_+ + h.c. \right) $ \\
        $ \hat{H}_{I,\text{eff}}^{(8)} \,=\, \frac{5}{\Lambda_{\Delta}^7} \hat{\sigma}_z $ 
        & $ \hat{\Phi}^{(8)} \,=\, 0 $ \\
    \end{tabular}
    \end{ruledtabular}
\end{table}

The conditions for convergence of the effective Hamiltonian for the far-detuned Rabi model have been obtained by Fel'dman \cite{feldmanConvergenceMagnusExpansion1984} and Fern\'{a}ndez \cite{fernandezConvergenceMagnusExpansion1990} using average Hamiltonian theory.
Here, repeating the recursive QAT algorithm \textit{ad nauseam} (see Table~\eqref{table:FarDetuned_Rabi_Problem}), we find that the QAT expansion also converges to the exact solution for $\lambda < \Lambda_\Delta$.
The coefficient of the $2k$th-order effective Hamiltonian term is $(-)^k C_{k}$ where $C_k$ is the $k$th Catalan number.
Since the generating function of the Catalan numbers is $c(x) = (1-\sqrt{1-4x})/(2x) = \sum_{k=0}^\infty C_k x^k$, we suspect by defining $x=-y^2$ for $y \equiv \lambda/\Lambda_\Delta$ and multiplying $c(x)$ by $\lambda y$, we will get this expansion. 
Indeed, we're looking at the series expansion of
\begin{equation}\label{eq::ex:Rabi_eff_Ham_ansatz}
\begin{split}
    \hat{H}_{I,\mathrm{eff}} ={}& -\!\lambda \, \frac{1-\sqrt{1+ (\sfrac{2\lambda}{\Lambda_\Delta})^2}}{(\sfrac{2\lambda}{\Lambda_\Delta})} \, \hat{\sigma}_z \\
    ={}& -\left(\frac{\Lambda_\Delta}{2} \pm \sqrt{\left( \frac{\Lambda_\Delta}{2}\right)^2 + \lambda^2} \right)\hat{\sigma}_z \\
    ={}& - \frac{1}{2\omega_{eg}} \left(\Delta \pm \sqrt{ \Delta^2 + \Omega^2} \right)\hat{\sigma}_z
\end{split}
\end{equation}
where the upper(lower) sign is for negative(positive) $\Delta$.
The dynamical phase expansion coincides with the series coefficients of $\arctan(x) = \sum_{n=0}^\infty \frac{(-1)^n}{2n+1} x^{2n+1}$ yielding
\begin{equation}\label{eq:ex:Rabi_phase_ansatz}
\hat{\Phi}(s;\lambda) = \frac{1}{2} \arctan{\left( \sfrac{2\lambda}{\Lambda_\Delta} \right)} \left(i \, e^{i \Lambda_\Delta s} \sigma_+ + h.c. \right)
\end{equation}
such that the fast propagator is
\begin{equation}\label{eq:ex:Rabi_QAT_ansatz}
	\begin{aligned}
		\hat{U}_\mathrm{fast}(s;\lambda) ={}& \exp{(-i \hat{\Phi}(s;\lambda))} \\
  \doteq{}&
		\begin{pmatrix}
			\cos{(\theta/2)} & -e^{i \Lambda_\Delta s} \sin{(\theta/2)}  \\
			e^{-i \Lambda_\Delta s} \sin{(\theta/2)} & \cos{(\theta/2)}  \\
		\end{pmatrix}
\end{aligned}
\end{equation}
where $\theta(\lambda) = \arctan{(2 \lambda / \Lambda_\Delta)} = \arctan{( \Omega / \Delta)}$ quantifies the strength of the perturbation over the principle branch $0 \le \theta \le \pi/2$. 
The form of the fast propagator implies that the effective dynamics are viewed from within a stationary frame rotated $\frac{\theta}{2}$ about the $z$-axis and precessing periodically with frequency $\Lambda_\Delta$.
Transforming the interaction Hamiltonian in eq.~\eqref{eq:ex:Rabi_interaction_hamiltonian} with respect to the fast propagator above returns a constant effective Hamiltonian in agreement with \eqref{eq::ex:Rabi_eff_Ham_ansatz} and the expansion terms in Table~\eqref{table:FarDetuned_Rabi_Problem}.
In appendix~\eqref{appendix:Rabi_problem} we show that the QAT solution is equivalent to \eqref{eq:ex:Rabi_full_solution} up to an irrelevant time-independent gauge transformation on the system.

\subsubsection*{Example 2: Resonant Two-Photon Raman Transition}\label{sec:qat:ex_Raman}

Consider two far-detuned, monochromatic laser tones interacting with a three-level system (3LS) where the qubit states $\ket{1}$ and $\ket{2}$ are coupled via an intermediate state $\ket{r}$ \cite{bergmannPerspectiveStimulatedRaman2015,sanzAdiabaticEliminationEffective2016}. 
The $\Lambda$-linkage pattern follows from two individually addressed 2LS with a shared excited state. 
The (dimensionless) interaction Hamiltonian is
\begin{equation}\label{eq:ex:Raman_interaction_Hamiltonian}
\hat{H}_I(s;\lambda) = \lambda \, \sum_{k=1,2} \left( \frac{\bar{\Omega}_k(s)}{2}  \ket{r}\bra{k} + h.c. \right)
\end{equation}
with normalized matrix elements
\begin{equation}\label{eq:ex:Raman_matrix_element}
    \bar{\Omega}_k(s) = 
\frac{\Omega_k}{\Omega_{\mathrm{max}}} e^{-i \Lambda_{\Delta_k} s} e^{i \phi_k}
\end{equation}
where $s = \omega_{0}^{(21)} t$ with the time variable $t$ scaled by the smallest energy scale of the unperturbed Hamiltonian, the transition frequency $\omega_{0}^{(21)}$ between states $\ket{1}$ and $\ket{2}$.
We define $\lambda = \Omega_{\mathrm{max}} / \omega_{0}^{(21)}$ with $\Omega_\mathrm{max} = \max(\{\Omega_k\})$ over the set of resonant Rabi frequencies $\Omega_k$, and $\Lambda_{\omega} \equiv \omega/\omega_{0}^{(21)}$ defines a dimensionless ``frequency'' for any true frequency $\omega$.
The detuning parameter $\Lambda_{\Delta_k} = \Lambda_{\omega_k} - \Lambda_{\omega_0}^{(rk)}$ is the difference between the drive $\Lambda_{\omega_k}$ and the energy level gap $\Lambda_{\omega_0}^{(rk)} > 0$ between states $\ket{k} \leftrightarrow \ket{r}$ for $k=1,2$.

We assume $|\Lambda_{\Delta_k}| \gg \lambda$, yet $\Lambda_{\Delta_1} = \Lambda_{\Delta_2} \equiv \Lambda_\Delta$ such that $\Lambda_{\Delta_1} - \Lambda_{\Delta_2} = (\Lambda_{\omega_1} - \Lambda_{\omega_2}) - \Lambda_{\omega_0}^{(21)} =0$ with the two laser beat note resonant with the transition frequency $\Lambda_{\omega_0}^{21}$ between states $\ket{1}$ and $\ket{2}$.
For the resonant Raman transition, the leading non-zero contribution are the second-order QAT results
\begin{subequations}
\begin{align}
		\hat{\Phi}^{(1)}(s;\lambda) =& \int^s ds' \; \hat{H}_I^{(1)}(s') \\
  \hat{H}^{(2)}_\mathrm{eff} ={}& \hat{H}_{\mathrm{R}} + \hat{H}_\mathrm{LS} \label{eq:ex:Raman_effective_Hamiltonian}
\end{align}
\end{subequations}
where, using the shorthand notation $\bar{\Omega}_k \equiv \bar{\Omega}_k(0)$,
\begin{equation}
    \hat{H}_\mathrm{R} = \frac{1}{4\Lambda_\Delta} \left( \bar{\Omega}_1 \bar\Omega_2^* \ket{2}\bra{1} + h.c. \right),
\end{equation}
which describes the process of a stimulated Raman transition in the $\Lambda$-linkage pattern for coherently transferring population between $\ket{1} \leftrightarrow \ket{2}$ despite no direct interaction.
The additional effective Hamiltonian contribution, $\hat{H}_\mathrm{LS} =\sum_{k=1,2} \hat{H}_{\mathrm{LS},k}$ where
\begin{equation}
\begin{aligned}
        \hat{H}_{\mathrm{LS},k} ={}& \frac{|\bar\Omega_k|^2}{4\Lambda_\Delta}  \left( \ket{k} \bra{k} - \ket{r}\bra{r} \right),
\end{aligned}
\end{equation}
generates energy level shifts due to the far-detuned interactions, yielding a differential shift on the qubit manifold for $|\bar\Omega_1| \neq |\bar\Omega_2|$.
The approximate second-order QAT dynamics for the interaction picture propagator governed by $\hat{H}_I(s;\lambda)$ are given by $\hat{U}_I^{[2]}(s) = \hat{U}_{\mathrm{fast}}^{[1]}(s;\lambda) \hat{U}_\mathrm{eff}^{[2]}(s;\lambda) = \exp(-i \hat{\Phi}^{[1]}(s;\lambda)) \exp(-i \hat{H}_\mathrm{eff}^{[2]}[\lambda] s)$ with bounded error estimate $\mathcal{O}(\lambda)$ valid at least over a time $s\sim \mathcal{O}(1/\lambda^2)$.

As in the far-detuned Rabi problem, we find the terms of the QAT expansion to be sufficiently simple as to recover an exact solution.
Using the recursive QAT algorithm, we find that the $(2k+2)$th-order effective Hamiltonian term is
\begin{equation}
    \hat{H}_\mathrm{eff}^{(2k+2)} ={} f_k \, \hat{H}_\mathrm{eff}^{(2)}, \quad k \ge 0
\end{equation}
where the series coefficient $f_k ={} (-1)^k \, C_{k} \bigl[ \bigl(|\bar\Omega_1|^2+|\bar\Omega_2|^2 \bigr)/4\Lambda_\Delta^{2}\bigr]^{k}$ and $C_k$ is the $k$th Catalan number.
From the generating function $c(x) = (1- \sqrt{1-4x})/2x=\sum_{k\ge0} C_k x^k$ with $x = -(\sfrac{\lambda}{2 \Lambda_\Delta})^2 \bigl(|\bar\Omega_1|^2+|\bar\Omega_2|^2 \bigr)$ we find the closed-form expression
\begin{equation}
\begin{aligned}
    \hat{H}_\mathrm{eff} ={}& \lambda^2 c(x)\hat{H}_\mathrm{eff}^{(2)} \\
    ={}& -\frac{2\Lambda_\Delta}{\bar\Omega_\mathrm{rms}^2}\left(\Lambda_\Delta\mp \sqrt{\Lambda_\Delta^2+\lambda^2 \bar\Omega_\mathrm{rms}^2}\right) \hat{H}_\mathrm{eff}^{(2)}
\end{aligned}
\end{equation}
where $\bar\Omega_\mathrm{rms}=\sqrt{|\bar\Omega_1|^2+|\bar\Omega_2|^2}$ and the upper (lower) sign corresponds to positive (negative) detuning $\Lambda_\Delta$.
Further, the $(2k+1)$th-order dynamical phase term is
\begin{equation}
    \hat{\Phi}^{(2k+1)}(s) ={} g_k \, \hat{\Phi}^{(1)}(s), \quad k \ge 0
\end{equation}
where $g_k = (-1)^k [ (|\bar\Omega_1|^2+|\bar\Omega_2|^2)/\Lambda_\Delta^{2}]^{k}/ (2k+1)$.
From the generating series $\arctan(x)/x = \sum_{n\ge0} \frac{(-1)^n}{2n+1} x^{2n}$ with $x = \lambda \bar\Omega_\mathrm{rms}/\Lambda_\Delta$ we find the closed-form expression
\begin{equation}
\begin{aligned}
    \hat{\Phi}(s) ={}& \frac{\Lambda_\Delta}{\bar\Omega_\mathrm{rms}} \arctan(\sfrac{\lambda \bar\Omega_\mathrm{rms}}{\Lambda_\Delta}) \, \hat{\Phi}^{(1)}(s).
\end{aligned}
\end{equation}
While equivalent up to a gauge transformation to the typical bright/dark dressed state solution \cite{bergmannPerspectiveStimulatedRaman2015}, the QAT solution explicitly captures the resonant coupling between the lower-lying qubit states in the effective Hamiltonian description with the off-resonant coupling to the intermediate state contained in the dynamical phase.

\begin{figure}[!tp]
    \centering
    % \includegraphics[width=8.6cm]{plots/Raman_plot.pdf}
    %\resizebox{1.01\columnwidth}{!}{
    \includestandalone[width=\columnwidth]{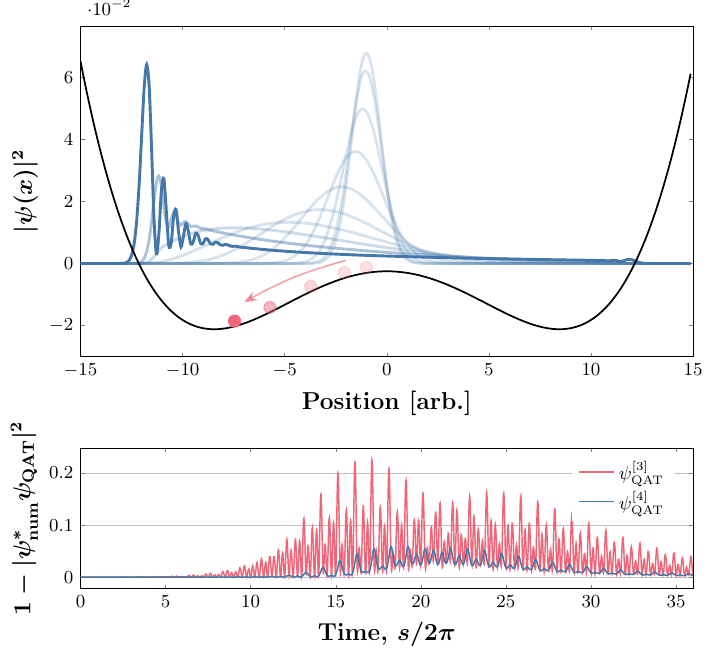}
    %}
    % \includegraphics[width=\textwidth]{plots/MS_Gate.pdf}
    %\caption{Population transfer by stimulated Raman transition in $\Lambda$-linkage pattern system. \KBComment{Initial state is $\ket{\psi(0)}=\ket{1}$. Give parameters for simulation. $\delta_1-\delta_2\neq0$ s.t. imperfect population transfer. Gray line is effective dynamics. dashed lines are numerical simulation. colored lines are analytic approximation.}}
    \caption{
    \textbf{Time-Evolution of the Particle Wavefunction in a Driven $1\mathrm{D}$ Double-Well Potential.} A quantum particle of unit mass begins in a minimum-uncertainty coherent state centered at $x = -1$ (arb. units) and initially at rest. 
    A weak, effective double-well potential arises due to the high-frequency modulation of the trap. 
    While the classical trajectory follows the gradient toward the nearest potential minimum, the anharmonicity induces quantum spreading and rapid loss of coherence in the wavefunction. To illustrate these dynamics, we choose: natural oscillator frequency $\omega_0 = 0.2\omega$, DC confinement $\omega_{dc} = 1.5\omega_\mathrm{eff}$ with $\omega_\mathrm{eff} = \omega_0^2 / 2\omega$, and anharmonic coupling strength $g = 1.75\cdot10^{-2}$. At leading order, the QAT approximation moderately agrees with the full numerical solution; including the next non-zero correction significantly improves accuracy in both amplitude and phase estimation.
    }
    \label{fig:DoubleWell_wavefcn}
\end{figure}
\begin{figure}[!tp]
    \centering
    % \includegraphics[width=8.6cm]{plots/Raman_plot.pdf}
    %\resizebox{1.01\columnwidth}{!}{
    \includestandalone[width=\columnwidth]{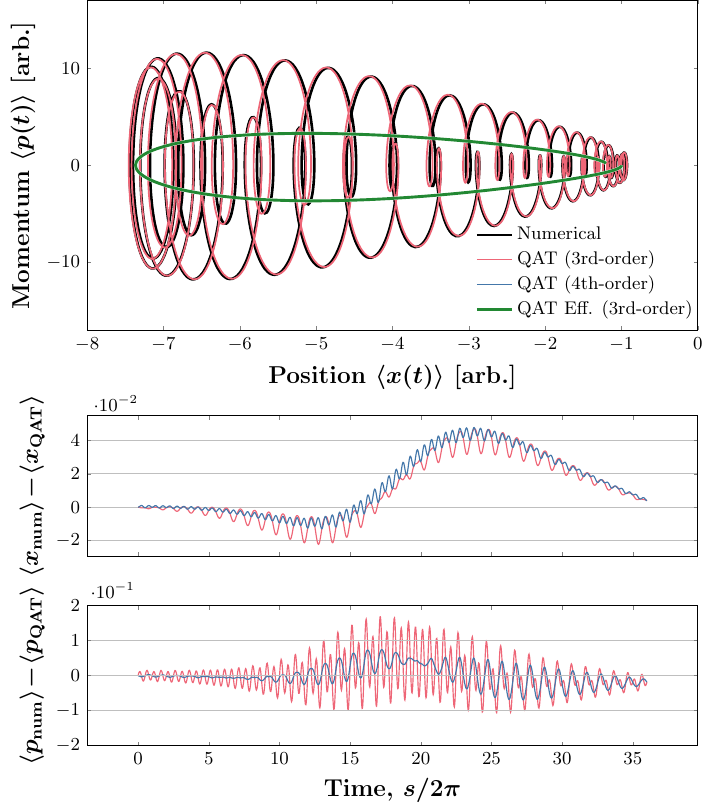}
    %}
    % \includegraphics[width=\textwidth]{plots/MS_Gate.pdf}
    \caption{
    \textbf{Phase Space of Particle in a Driven $1\mathrm{D}$ Double-Well Potential.} Using the same parameters as in Fig. 2, we plot the phase space trajectory of the particle. 
    The high-frequency modulation gives rise to an effective time-averaged trapping potential, which generates a ponderomotive force. 
    Under these conditions, the particle exhibits motion centered around the effective secular trajectory but exhibits pronounced excess micromotion due to the relatively large ratio $\omega_0/\omega_\mathrm{rf} = 0.2$.
    }
    \label{fig:DoubleWell_xp}
\end{figure}

\subsubsection*{Example 3: Time-Averaged Double-Well Potential}\label{sec:qat:ex_AnharmonicTrapping}
We now consider a single quantum particle subject to harmonic driving and weak anharmonic confinement potentially realizable, for example, on a trapped-ion QCCD architecture \cite{bowlerCoherentDiabaticIon2012,qinCharacterizingSpatialPotential2025} or in a double-well quantum dot \cite{pettaManipulationSingleCharge2004,zhengMinimalDoubleQuantum2017}. 
For suitable choices of the drive amplitudes and frequency, the resulting time-averaged potential forms an effective weak double-well structure centered near the trap origin.
The system Hamiltonian is given by
\begin{equation}
    \begin{split}
    \hat{H}_\mathrm{S}(t)=\frac{\hat{p}^2}{2m} + \frac{1}{2}m\omega_0^2 \hat{x}^2 \cos(\omega_\mathrm{rf} t)  \\
    - \frac{1}{2} m \omega_\mathrm{dc}^2 \hat{x}^2+\frac{1}{4}g \hat{x}^4
    \end{split}
\end{equation}
where $\omega_0$ is the natural oscillator frequency, $\omega_\mathrm{dc}$ characterizes the DC potential with $\omega_\mathrm{dc} \ll \omega_0$, $\omega_\mathrm{rf}$ is the rf-driving frequency, and $g= m \omega_0^2/l_\mathrm{g}^2$ is the anharmonic coupling strength, which can be expressed in terms of a characteristic length scale of the anharmonicity $l_\mathrm{g}$.
% The parameter $q = \pm 1$ sets the sign of the DC potential, which determines whether we have a double-well potential (negative) or a purely anharmonic trap (positive).
Expressing the Hamiltonian in dimensionless terms we have the time coordinate $s=\omega_\mathrm{rf}t$, position $\hat{x} = l_\mathrm{o} \hat{x}^\prime$, and momentum $\hat{p} = \hbar \hat{p}^\prime/l_\mathrm{o}$ for $l_\mathrm{o} = \sqrt\frac{\hbar}{m\omega_0}$ yields
\begin{equation}
\begin{split}
    \hat{H}_\mathrm{S}(s) = \lambda \hat{H}_\mathrm{S}^{(1)}+\lambda^3\hat{H}_\mathrm{S}^{(3)}
\end{split}
\end{equation}
where
\begin{subequations}
    \begin{align}
        \hat{H}_\mathrm{S}^{(1)} ={}&  \frac{\hat{p}^{\prime 2}}{2} + \frac{1}{2} \hat{x}^{\prime 2} \cos(s) \\
        \hat{H}_\mathrm{S}^{(3)} ={}& \left(\frac{\omega_\mathrm{rf}}{\omega_0} \right)^2 \left[ - \frac{1}{2} \frac{\omega_\mathrm{dc}^2}{\omega_0^2} \hat{x}^{\prime 2} + \frac{1}{4}  \frac{l_\mathrm{o}^2}{l_\mathrm{g}^2}  \hat{x}^{\prime 4} \right]
    \end{align}
\end{subequations}
where $\lambda = \sfrac{\omega_0}{\omega_\mathrm{rf}}\ll 1$ and $\sfrac{\omega_\mathrm{dc}}{\omega_0} \sim \sfrac{l_\mathrm{o}}{l_\mathrm{g}} \sim\mathcal{O}(\lambda)$.
The dimensionless position $\hat{x}^\prime$ and momentum $\hat{p}^\prime$ have the canonical commutation relation $[\hat{x}^\prime, \hat{p}^\prime] = i$.
Typically, one would transform into an interaction picture toggling at the rf-frequency; however, we will not transform the system in this example.
In this case, the validity of the result will depend on each term in the perturbative QAT expansion becoming increasingly smaller, which is satisfied by the expansion in the high-frequency drive.
To lowest order, the effective dynamics are of a slowly-moving free particle given by the time-averaged dynamics $\hat{H}_\mathrm{eff}^{(1)} = \timeavg{\hat{H}_\mathrm{S}^{(1)}}{s} = \hat{p}^{\prime 2}/2$.
The more interesting trapping dynamics are found at the next non-zero effective Hamiltonian contribution given by the third-order QAT result
\begin{subequations}
    \begin{align}
        \lambda^3 \hat{H}_\mathrm{eff}^{(3)} ={}& \frac{1}{2}\left( \frac{\omega_\mathrm{eff}^2}{\omega_0\omega_\mathrm{rf}} - \lambda \frac{\omega_\mathrm{dc}^2}{\omega_0^2} \right) \hat{x}^{\prime 2} + \frac{\lambda}{4}  \frac{l_\mathrm{o}^2}{l_\mathrm{g}^2} \hat{x}^{\prime 4} 
    \end{align}
\end{subequations}
and
\begin{subequations}
    \begin{align}
        \lambda \hat{\Phi}^{(1)}(s) ={}& \frac{\lambda}{2} \sin(s) \hat{x}^{\prime 2} \\
        \lambda^2 \hat{\Phi}^{(2)}(s) ={}& \frac{\lambda^2}{2} \cos(s) \left\{\hat{x}^\prime, \hat{p}^\prime\right\} 
    \end{align}
\end{subequations}
where $\{\cdot,\cdot\}$ is the anti-commutator and we have defined the effective secular frequency in the usual form $\omega_\mathrm{eff} = \omega_0^2/\sqrt{2}\omega_\mathrm{rf}$ for a purely harmonic trapping potential.
In dimensionful units we have 
\begin{equation}
    \hat{H}_\mathrm{eff}^{[3]} = \frac{\hat{p}^2}{2m} + \frac{1}{2}m(\omega_\mathrm{eff}^2-\omega_\mathrm{dc}^2) \hat{x}^2 + \frac{1}{4{l_\mathrm{g}^2}} m \omega_0^2 \hat{x}^4
\end{equation}
where $g=0$ (i.e. $l_g \rightarrow \infty$) returns an (inverted) harmonic trapping potential for $\omega_\mathrm{eff} > \omega_\mathrm{dc}$ ($\omega_\mathrm{dc} > \omega_\mathrm{eff}$) as expected.
For this simple example we find that the next non-zero correction to the effective Hamiltonian does not occurs until seventh order.
The time-evolution of a particle's wavefunction and phase space in the effective double-well potential are shown in Fig.~\eqref{fig:DoubleWell_wavefcn} and ~\eqref{fig:DoubleWell_xp}, respectively.

\section{Quantum Averaging on Multiple Timescales}\label{sec:qat_multiplescales}

Perturbations involving near-resonant Fourier modes or higher-order mode interactions pose a unique challenge to maintaining asymptotic validity, which we have so far ignored. 
Yet, consider the Rabi problem in Sec.~\eqref{sec:qat:ex_Rabi} now with a real-valued AC field, $\hat{V}(t)/\hbar = \Omega \cos(\omega t) \hat{\sigma}_x$ for which no closed-form solution is known \cite{shirleySolutionSchrodingerEquation1965}.
When driven near resonance, the system exhibits both slow population inversion known as Rabi flopping overlaid with fast beat-note dynamics, exemplifying the distinct behavior occurring on different timescales that demands more sophisticated modeling approaches.

One widely used method for simplifying multi-timescale dynamics is the rotating-wave approximation (RWA), which separates fast and slow modes in the interaction Hamiltonian set by a high-frequency cutoff and eliminates the former \cite{sakuraiModernQuantumMechanics2020}. 
Specifically, the interaction Hamiltonian is decomposed into high- and low-frequency components, $\hat{H}_I^{>}(s)$ and $\hat{H}_I^{<}(s)$, based on a cutoff frequency $\Lambda_\mathrm{c}$.
The RWA then effectively acts as an idealized low-pass filter to eliminate rapidly oscillating counter-rotating terms.
The long-time features of the system are then approximated by the coarse-grained Hamiltonian \begin{equation}\label{eq:qat_multiplescales:coarse_grained_perturbation} \hat{H}_{I}(s;\lambda) \simeq \hat{H}_{I,0}(\lambda)+ \hat{H}_I^{<}(s;\lambda) + \cancel{\hat{H}_I^{>}(s;\lambda)}. \end{equation}
While simple, the RWA fails to capture higher-order interactions and subtle near-resonant effects.
For example, if $\hat{H}_I^<(s;\lambda) = 0$ but nearly-resonant beat notes arise from higher-order interactions, the first-order approximation vanishes without a clear path for corrections. 
Yet, such interactions can yield significant physical phenomena, as seen in two-photon Raman transitions \cite{bergmannPerspectiveStimulatedRaman2015} or Mølmer-Sørensen entanglement interactions \cite{sorensenQuantumComputationIons1999}. 
These limitations have motivated numerous techniques to gain analytic insight beyond the RWA \cite{frascaUnifiedTheoryQuantum1992,yanEffectsCounterrotatingCouplings2017,kohlerDispersiveReadoutUniversal2018,zeuchExactRotatingWave2020,wangQuantumNonreciprocalInteractions2023} and the closely related adiabatic elimination method \cite{paulischAdiabaticEliminationHierarchy2014,sanzAdiabaticEliminationEffective2016,bergmannPerspectiveStimulatedRaman2015,ribeiroSystematicMagnusBasedApproach2017}.
Building on these methods, we aim to develop a unified framework that systematically improves effective quantum models while addressing the full complexity of multi-timescale dynamics.

To address this challenge, we propose a multiple timescale analysis (MTSA) to systematically separate fast-varying dynamics from slow, adiabatic behavior \cite{mitropolskiiProblemsAsymptoticTheory1965,kevorkianPerturbationTechniquesOscillatory1987,nayfehMethodMultipleScales2000}.
We begin by introducing a two-timescale derivative expansion, establishing a rigorous framework for analyzing multi-timescale dynamics.
Building on this foundation, we present a generalized QAT framework utilizing a Partitioned Expansion by Timescale Separation (PETS) approach, an intuitive RWA-based regularization technique designed to separate timescales, as detailed in the companion paper \cite{barajasMultiTimescaleCoherentControl2025}.
We demonstrate that the PETS approach, an inductive method, and the two-timescale derivative expansion, a deductive method, provide consistent two-timescale QAT frameworks.
Finally, we extend the two-timescale renormalization procedure to address multiple timescale dynamics through an iterative, Multiple PETS procedure.

\subsection{Two-Timescale Derivative Expansion Framework}\label{sec:qat_multiplescales:twotime}

In a derivative-based MTSA approach we introduce two disparate ``time'' variables, that although not uncorrelated are operationally treated as such.
That is, when $\hat{H}_I(s;\lambda)$ generates interactions on two different timescales, we assume all time-dependent operators have the form
\begin{equation}
    \hat{A}(s) \equiv \hat{A}(s_0,s_1)
\end{equation}
for time variables $s_i$ regarded as independent of each other.
From previous discussions we have already identified (at least) two timescale regimes of the system: the rapidly varying $s \lesssim \mathcal{O}(1)$ dynamics modulating the slowly varying $\Lambda_\mathrm{c} s \gtrsim \mathcal{O}(1)$ dynamics set by a high-frequency cutoff $\Lambda_\mathrm{c} \ge \lambda$.
This behavior is exemplified in the resonantly driven Rabi problem, where the minimum time required for coherent state transfer is inversely proportional to the Rabi frequency, which is determined by the driving amplitude of the perturbation.
If we define a fast timescale $\sigma \equiv s_0 = s$ and a ``lagging'' slow timescale $\tau \equiv s_1 = \Lambda_\mathrm{c} s$, from the chain-rule we have the two-timescale derivative expansion \cite{nayfehMethodMultipleScales2000}
\begin{equation}\label{eq:qat_multiplescales:twotime_derivative}
\frac{d}{ds} \doteq \left(\frac{\partial}{\partial \sigma}\right)_{\tau} + \Lambda_\mathrm{c} \left(\frac{\partial}{\partial \tau}\right)_{\sigma},
\end{equation}
which is an exact procedure that becomes increasingly tractable as the separation increase (i.e. $\lambda \le \Lambda_\mathrm{c} \ll 1$).
Replacing the time-derivative in the homological equation \eqref{eq:qat:QAT_homological} with \eqref{eq:qat_multiplescales:twotime_derivative} and using $\Lambda_\mathrm{c} = \lambda \, \Lambda_\mathrm{c}^{<}$ confirms a consistent asymptotic expansion is obtained if $\lambda \le \Lambda_\mathrm{c} <1$.
Only at the end of the calculation do we return to the true time variable $s$ by appropriate substitution.

Inspired by Lindstedt–Poincaré method \cite{nayfehMethodMultipleScales2000}, we introduce a ``frequency decomposition'' method to parameterize $\hat{H}_I(s) \equiv \hat{H}_I(\sigma,\tau)$ explicitly in terms of the two timescales.
The frequency decomposition is the process of parameterizing the frequency modes $\Lambda_{\omega_k}$ of the multi-modal $\hat{H}_I(s)$ in powers of $\Lambda_\mathrm{c}$ in terms of unknown regularization parameters.
The regularization parameters will be determined to ensure a valid two-timescale asymptotic expansion and when resummed must equal the bare mode frequency.
Explicitly, let each base frequency $\Lambda_{\omega_k} \in \vec{\Lambda}$ be expanded in powers of $\Lambda_\mathrm{c}$ as follows:
\begin{equation}\label{eq:qat_multiplescales:freq_decomposition}
\begin{aligned}
    \Lambda_{\omega_k} ={}& \sum_{n \ge0} \Lambda_\mathrm{c}^n \Lambda_{\omega_k}^{(n)} \\
    ={}& \Lambda_{\omega_k}^{>} + \Lambda_\mathrm{c} \, \Lambda_{\omega_k} ^{<}(\Lambda_\mathrm{c}),
\end{aligned}
\end{equation}
where $\Lambda_{\omega_k}^{(n)}$ are the frequency regularization parameters and we define a high-frequency contribution $\Lambda_{\omega_k}^{>} = \Lambda_{\omega_k}^{(0)}$ and the rescaled low-frequency contribution $\Lambda_{\omega_k}^{<} \equiv \sum_{n\ge0} \Lambda_\mathrm{c}^n \Lambda_{\omega_k}^{(n+1)} \le 1$.
From eq.~\eqref{eq:qat_multiplescales:freq_decomposition}, it follows that the generalized phase of a mode may be expressed as
\begin{equation}\label{eq:qat_multiplescales:Fourier_phase}
\begin{aligned}
	\Lambda_{\omega_k} s ={}& \bigl( \Lambda_{\omega_k}^{>} + \Lambda_\mathrm{c} \, \Lambda_{\omega_k} ^{<} \bigr) s \\
    ={}& \Lambda_{\omega_k}^{>} \sigma + \Lambda_{\omega_k}^{<} \tau,
\end{aligned}
\end{equation}
where we have identified the two timescales as desired.
We remark that in general the regularization parameters are not uniquely determined. 
Regularization parameters valid to first-order approximation are given by
\begin{equation}
    \Lambda_{\omega_k} = 
    \begin{cases}
        \Lambda_{\omega_k}^>, \quad &\text{if } \Lambda_{\omega_k} > \Lambda_c \\
        \lambda \Lambda_{\omega_k}^<, \quad &\text{otherwise} \\
    \end{cases}
\end{equation}
A valid $N$th-order approximation for $N \ge 2$ requires identifying regularization parameters $\Lambda_{\omega_k}^{(n)}$ for $0\le n \le N-1$ for each $k$th-mode, which can be entirely deduced from the procedure in Table~\eqref{table:twotime_freq}.
%%% Two-Timescale Table %%%%%%%%%%%%%%%%%%%%%%%%
\begin{table}[t]
\caption{\textbf{Frequency Decomposition Procedure.} The decomposition process for near-resonant base frequency (first line) and far-detuned base frequency (second line).
The latter may interact near-resonantly with other modes at higher orders.
In the following, $\vec{\alpha} \in \mathbb{Z}^{\mathrm{dim}(\vec{\Lambda})}$ is an arbitrary integer vector with $|\vec{\alpha}|\neq0$.
}
\label{table:twotime_freq}
\begin{ruledtabular}
\setlength{\tabcolsep}{6pt} % Default value: 6pt
\begin{tabular}{c@{}l}
Frequency & Decomposition \\
\hline

% == Row 1 == %
\makecell{$\Lambda_{\omega_k} \le \Lambda_c$}
&
\makecell{$\Lambda_{\omega_k} \equiv \lambda \Lambda_{\omega_k}^<$} \\
\hline

% % == Row Two == %
% \makecell{
%     $\Lambda_{\omega_k} > \Lambda_c$ and $\displaystyle |\vec{\alpha} \cdot \vec{\Lambda}| > \Lambda_c$ \vspace{0.75em} \\
%     for \textit{all} $\vec{\alpha}$ with $\alpha_k \Lambda_{\omega_k} \neq 0$
%     % $\Lambda_{\omega_k} > \Lambda_c$ and \vspace{0.75em} \\
%     % $\forall \alpha_j \in \mathbb{Z}$ and $\alpha_k \neq 0$, \\
%     % $\displaystyle \sum_{\mathclap{\Lambda_{\omega_j} \! \in \vec{\Lambda}}} \alpha_j \Lambda_{\omega_j} > \Lambda_c$
% } &

% \makecell{$\Lambda_{\omega_k} \equiv \Lambda_{\omega_k}^{>}$ \\
% } \\
% \hline

% == Row Three == %
\makecell{
    $\Lambda_{\omega_k} > \Lambda_c$ 
    % and $\displaystyle |\vec{\alpha} \cdot \vec{\Lambda}| \le \Lambda_c$ \vspace{0.75em}\\
    % for \textit{any} $\vec{\alpha}$ with $\alpha_k \Lambda_{\omega_k} \neq 0$ 
} &
\makecell[cl]{To $N$th-order approximation: \\
($1$) Define $\Lambda_{\vec{\alpha}} =\vec{\alpha} \cdot \vec{\Lambda}$ for \\
each $\vec{\alpha}$ where $|\vec{\alpha}|\le N$ and $\alpha_k \neq 0$. \\
($2$) Use eq.~\eqref{eq:qat_multiplescales:freq_decomposition} to equate parameters \\
 in powers of $\Lambda_c$ up to $\mathcal{O}(\Lambda_c^{N})$. \\
($3$) From system of equations, \\
determine parameters.} \\
% \, \Lambda_{\vec{\alpha}}^<=\vec{\alpha} \cdot \vec{\Lambda}
\end{tabular}
\end{ruledtabular}
\end{table}

The two-timescale QAT formalism largely mirrors the results from \cref{sec:qat} with minor modifications to accommodate the two-timescale derivative expansion.
Unlike before, the effective Hamiltonian will now depend on the ($\tau$) slow-time and govern the long-time dynamics from $\hat{H}_I(s;\lambda)$ free of the fast-time degree of freedom.
For the two-timescale derivative expansion the QAT homological equation becomes
\begin{multline}\label{eq:qat_multiplescales:QAT_homological}
    \partial_\sigma \hat{\Phi}^{(n)}(\sigma,\tau) = \hat{\mathcal{H}}^{(n)}_{\Phi}(\sigma,\tau) - \hat{H}_{I,\mathrm{eff}}^{(n)}(\tau) \\
    -  \Lambda_\mathrm{c}^{<} \partial_{\tau}\hat{\Phi}^{(n-1)}(\sigma,\tau)
\end{multline}
with the two-timescale solution
\begin{multline}\label{eq:qat_multiplescales:QAT_homological_integral}
	\hat{\Phi}^{(n)}(\sigma,\tau) = \int^\sigma d\sigma' \left( \hat{\mathcal{H}}_{\Phi}^{(n)}(\sigma',\tau) - \hat{H}_{I,\mathrm{eff}}^{(n)}(\tau) \right. \\
 - \left. \Lambda_\mathrm{c}^{<} \partial_{\tau}\hat{\Phi}^{(n-1)}(\sigma',\tau) \right),
\end{multline}
where we've enforced the same uniqueness rule as before.
Consider that \eqref{eq:qat_multiplescales:QAT_homological_integral} implies adiabatic passage with respect to $\tau$-time dependence.
The additional term on the second line corresponds to an adiabatic correction from the previous order that ensures the procedure remains exact.

To regularize with respect to the fast $\sigma$-time we adapt the time-averaging procedure to two timescales with
\begin{equation}\label{eq:qat_multiplescales:QAT_time_average}
	\timeavg{\hat{A}^{(n)}(s)}{\sigma} = \lim_{T \to \infty} \frac{1}{T} \int_{{\sigma_0}}^{{T+\sigma_0}} {\hat{A}^{(n)}(\sigma',\tau)} \: d{\sigma'} {},
\end{equation}
holding the $\tau$-time constant.
Clearly, the two-timescale averaging procedure is effectively applying a low-pass filter on high-frequency components in the same manner as in the RWA. 
We are now able to systematically ``integrate'' out only fast-time effects allowing us to adiabatically follow the slowly-varying $\tau$-time interactions.
It becomes clear that the expansion is regularized if
\begin{equation}\label{eq:qat_multiplescales:QAT_phase_cond}
	\timeavg{\hat{\Phi}^{(n)}(\sigma,\tau)}{\sigma} = 0,
\end{equation}
which is satisfied for the $\tau$-time dependent effective Hamiltonian
\begin{equation}\label{eq:qat_multiplescales:QAT_effHam_cond}
	\hat{H}_{I,\mathrm{eff}}^{(n)}(\tau) = \timeavg{\hat{\mathcal{H}}_{\Phi}^{(n)}(\sigma,\tau)}{\sigma} = \hat{\mathcal{H}}_{\Phi,0}^{(n)} + \hat{\mathcal{H}}_{\Phi,<}^{(n)}(\tau)
\end{equation}
governing resonant \textit{and} near resonant interactions.
Accounting for these changes, it has been shown that the main results of Sec.~\eqref{sec:qat} still apply to two-timescale expansions \cite{perkoHigherOrderAveraging1969,kevorkianPerturbationTechniquesOscillatory1987}.
If the $\tau$-dependent effective Hamiltonian can't be explicitly solved, the effective propagator may be approximated with exponential perturbation theory (see  appendix \eqref{appendix:effective_equation_PT} for details). 

\subsubsection*{Example 1: Near-Resonant Rabi Problem with Two-Timescale Derivative Expansion}\label{sec:qat_multiplescales:Rabi_twotime}

The RWA Rabi problem studied in the far-detuned limit in \cref{sec:qat:ex_Rabi} demonstrated the basic features of the formalism with a single fast timescale. 
We now turn to the Rabi problem for the 2LS near-resonantly interacting with a real-valued AC field for which no known explicit solution exists: 
\begin{equation}\label{eq:ex:Rabi_CR_total_Hamiltonian}
	\hat{H}_\mathrm{total}(t) = \frac{\omega_{eg}}{2} \, \hat{\sigma}_z + \Omega \, \cos(\omega t) \, \hat{\sigma}_x
\end{equation}
where the driving frequency $\omega > 0$ and the Pauli x-operator $\hat{\sigma}_x = \hat{\sigma}_+ + \hat{\sigma}_-$. 
Expanding $\hat{H}_\mathrm{total}(t)$ in exponentials reveals the additional ``counter-rotating'' terms, $e^{\pm i \omega t} \hat{\sigma}_{\pm}$, not apparent in \eqref{eq:ex:Rabi_total_Hamiltonian} that are highly detuned in the interaction picture.
These terms generate fast-varying interactions and are dropped in the rotating-wave approximation. 

In the perturbative limit ($\omega_{eg} \gg \Omega$) we define the small parameter $\lambda = \Omega/2\omega_{eg}$ using the same convention as before. 
As previously mentioned, the choice of $\hat{H}_0(t)$ is somewhat arbitrary but often field-dependent. 
In this case two reasonable choices exist: the first is $\hat{H}_0 = \frac{\omega_{eg}}{2} \, \hat{\sigma}_z$ and the second is the RWA Rabi Hamiltonian whose solution is given by \eqref{eq:ex:Rabi_full_solution}. 
We will use the former as it is a more standard choice, but the latter is a compelling option if one is interested in studying the perturbative effects of the additional counter-rotating terms.
Before proceeding we rescale the system with the dimensionless time $s = \omega_{eg} t$ yielding $\hat{H}_0 = \hat{\sigma}_z / 2$ and $\hat{V}(s;\lambda) = \lambda \hat{V}^{(1)}(s)$ for $\hat{V}^{(1)}(s) = e^{-i \Lambda_\omega s} \, \hat{\sigma}_x + h.c.$ where $\Lambda_\omega = \omega/\omega_{eg}$.

\subparagraph{Interaction Picture.} In the interaction picture of $\hat{H}_0 = \hat{\sigma}_z/2$ we have
\begin{equation}
	\hat{H}_I^{(1)}(s) = \left(e^{-i \Lambda_\Delta s} + e^{i \Lambda_\Sigma s} \right) \hat{\sigma}_+ + h.c.
\end{equation}
where $\Lambda_\Delta =  \Lambda_\omega - \Lambda_{\omega_{eg}}$ is the detuning parameter and $\Lambda_\Sigma = \Lambda_{\omega_{eg}} + \Lambda_\omega = 2 + \Lambda_\Delta$ is the frequency of the counter-rotating terms satisfying $\Lambda_\Sigma \gg  \Lambda_\Delta$ for any driving frequency $\Lambda_\omega > 0$.

\subparagraph{Frequency Decomposition.} Suppose the detuning parameter is near resonance (i.e. $|\Lambda_\Delta| \le \Lambda_\mathrm{c}$) such that $|\Lambda_\Sigma| >
2 \gg \Lambda_\mathrm{c}$. 
Using Table~\eqref{table:twotime_freq} we identify the slow frequency $\Lambda_\Delta \equiv \lambda \Lambda_\Delta^{<}$ and the fast frequency $\Lambda_\Sigma \equiv \Lambda_\Sigma^{>}$, a decomposition valid up to $(n+m)$-th order where the mode interactions generate the nearly resonant beat note $n \Lambda_\Sigma - m \Lambda_\Delta \le \lambda$.
From eq.~\eqref{eq:qat_multiplescales:Fourier_phase} we have the two-time interaction Hamiltonian
\begin{equation}
	\hat{H}_I^{(1)}(\sigma,\tau) = \left( e^{-i \Lambda_\Delta^{<} \tau} + e^{i \Lambda_\Sigma \sigma} \right) \hat{\sigma}_+ + h.c.
\end{equation}

\subparagraph{QAT Homological Equation.} Let $\lambda$ be sufficiently small such that the system is well-characterized by its second-order dynamics.
For $n=1$, from Table~\ref{table:qat:auxiliary_hamiltonian} for $\hat{\mathcal{H}}_{\Phi}^{(n)}(s)$ and the two-time averaging procedure in eq.~\eqref{eq:qat_multiplescales:QAT_time_average},
\begin{equation}
    \begin{aligned}
            \hat{\mathcal{H}}_{\Phi}^{(1)}(s) ={}& \hat{H}_I^{(1)}(s) \\
	    \timeavg{\hat{\mathcal{H}}_{\Phi}^{(1)}(s)}{\sigma} ={}& \lim_{T \to \infty} \frac{1}{T}\int_{{0}}^{{T}} {\hat{H}_I^{(1)}(\sigma,\tau)} \: d{\sigma} \\
     ={}&  e^{-i \Lambda_\Delta^{<} \tau} \, \hat{\sigma}_+ + h.c.
     \end{aligned}
\end{equation}
leading to the first-order contributions
\begin{subequations}
	\begin{align}
        \hat{H}_{I,\mathrm{eff}}^{(1)}(\tau) ={}& \timeavg{\hat{\mathcal{H}}_{\Phi}^{(1)}(s)}{\sigma} = e^{-i \Lambda_\Delta^{<} \tau} \, \hat{\sigma}_+ + h.c. \label{eq:ex:twotime_Rabi_QAT_eff_o1} \\
        \begin{split} \label{eq:ex:twotime_Rabi_QAT_dynphase_o1}
		\hat{\Phi}^{(1)}(s) ={}&  \int^\sigma \! d{\sigma'} \: {\left(\hat{\mathcal{H}}_{\Phi}^{(1)}(\sigma',\tau) - \hat{H}_{I,\mathrm{eff}}^{(1)}(\tau)\right)} \\
        ={}& \frac{1}{i \Lambda_\Sigma} e^{i \Lambda_\Sigma \sigma} \, \hat{\sigma}_+ + h.c.
        \end{split}
	\end{align}
\end{subequations}
Repeating the algorithm to next two order we have
\begin{subequations}
	\begin{align}
        \hat{H}_{I,\mathrm{eff}}^{(2)} ={}& \timeavg{\hat{\mathcal{H}}_{\Phi}^{(2)}(s)}{\sigma} = \frac{1}{\Lambda_\Sigma} \, \hat{\sigma}_z \\
        \begin{split}
	\hat{\Phi}^{(2)}(s) ={}& \int^\sigma \! d{\sigma'} \: \left(\hat{\mathcal{H}}_{\hat{\Phi}}^{(2)}(\sigma',\tau) - \hat{H}_{I,\mathrm{eff}}^{(2)} \right. \\
    &\hspace{5em} \left. - \partial_\tau \hat{\Phi}^{(1)}(\sigma',\tau) \right) \\
        ={}& \frac{2}{\Lambda_\Sigma^2} \sin(\Lambda_\Delta^{<} \tau + \Lambda_\Sigma \sigma) \, \hat{\sigma}_z
        \end{split}
	\end{align}
\end{subequations}
and
\begin{subequations}
	\begin{align}
        \hat{H}_{I,\mathrm{eff}}^{(3)} ={}& -\frac{1}{\Lambda_\Sigma^2}e^{-i \Lambda_\Delta^{<} \tau} \, \hat{\sigma}_+ + h.c.  \\
        \begin{split}
	\hat{\Phi}^{(3)}(s) ={}& -\frac{2 \Lambda_\Delta^<}{\Lambda_\Sigma^3} \sin(\Lambda_\Delta^{<} \tau + \Lambda_\Sigma \sigma) \, \hat{\sigma}_z  \\
    &{} +  \frac{1}{6 \Lambda_\Sigma^3} \Bigl[ -i \, \hat{\sigma}_{+} e^{-i \Lambda_\Delta^{<} \tau} \Bigl(8 \cos (\Delta  \tau +\Sigma  t)   \\
    &{} + 8 e^{-i(\Delta  \tau +\Sigma  t)} +3 e^{2i (\Delta  \tau +\Sigma  t)}  \Bigr) + h.c. \Bigr].
        \end{split}
	\end{align}
\end{subequations}
As expected, the first-order approximation returns the interaction Hamiltonian in the rotating-wave approximation, which captures the most significant long-time effects governing the system dynamics.
The lowest-order fast-time contribution $\hat{\Phi}^{(1)}(\sigma)$ shows a weak coupling to the counter-rotating terms, which off-resonantly drives the transition.
The second-order effective interaction, also due to the counter-rotating terms, produces the familiar Bloch-Siegert shift (or light shift).
Finally, the third-order dynamical phase contribution introduces the first adiabatic correction for assuming that $\Lambda_\Delta t = \Lambda_{\Delta}^{<} \tau$ is approximately constant under fast-time integration (i.e. $\partial_\tau \hat{\Phi}^{(2)}(\sigma,\tau) \neq 0 $).
We remind the reader that for an $N$th-order approximation, only the $(N-1)$th dynamical phase contribution is required.

\subparagraph{The Effective Interaction Picture.}
The second-order truncated effective Hamiltonian is given by
\begin{equation}\label{eq:ex:twotime_Rabi_effHam}
\begin{split}
	\hat{H}^{[2]}_{I,\mathrm{eff}}(\tau;\lambda) =&{} \sum_{n=1}^{2} \lambda ^{n} \, \hat{H}^{(n)}_{I,\mathrm{eff}}(\tau) \\
    =&{} \underbracket{\vphantom{\frac{1^2}{1_2}}\lambda \, (e^{-i \Lambda_\Delta^{<} \tau} \, \hat{\sigma}_+ + h.c.)}_{\text{RWA}} \, + \!\! \underbracket{\frac{\lambda^2}{\Lambda_\Sigma} \, \hat{\sigma}_z.}_{\text{resonance shift}}
\end{split}
\end{equation}
Consider that by a simple redefinition of parameters $\hat{H}^{[2]}_{I,\mathrm{eff}}(\tau;\lambda)$ can be mapped onto the exactly-solvable Rabi problem in \eqref{eq:ex:Rabi_total_Hamiltonian}.
Therefore, the exact solution to the truncated effective interaction picture equation is
\begin{equation}\label{eq:ex:Rabi_CR_Ueff}
		\hat{U}_\mathrm{eff}^{[2]}(\tau;\lambda) = e^{-i \Lambda_\Delta^{<} \hat{\sigma}_z \tau/2} \, e^{-i \, \left(-\tilde{\Lambda}_\Delta^{<} \, \hat{\sigma}_z + 2\hat{\sigma}_x \right) \tau/2 }
\end{equation}
where $\tilde{\Lambda}_\Delta^{<} = \Lambda_\Delta^{<} - \frac{2 \lambda}{\Lambda_\Sigma}$  is the resonance-shifted detuning parameter.

For comparison we proceed to solve the truncated problem with perturbation theory (see appendix~\eqref{appendix:effective_equation_PT}).
Let $\hat{U}_{\mathrm{eff},0}^\prime(\tau) = e^{-i \Lambda_\Delta^{<} \hat{\sigma}_z \tau/2}$$e^{-i \, \left(-\Lambda_\Delta^{<} \, \hat{\sigma}_z + 2\hat{\sigma}_x \right) \tau/2 }$ be the solution to the first-order, RWA dynamics.
Then, in the interaction picture of $\hat{H}_{I,\mathrm{eff}}^{(1)}(\tau)$ we have the new slow-time interaction Hamiltonian satisfying eq.~\eqref{eq:appendix:effective_equation_PT:twotime_perturbation_problem_standard_form},
\begin{equation}
\begin{aligned}
    \hat{H}_{I,\mathrm{eff}}^{\prime (2)}(\tau) ={}& \frac{1}{\Lambda_\Sigma^{} {\Lambda_{\Omega^\prime}^<}^2} \Bigl\{ \bigl(\Lambda_\Delta^{< \, 2} + 4 \cos(\Lambda_{\Omega^\prime}^< \tau) \bigr) \hat{\sigma}_z  \\
    & - (g(\Lambda_{\Omega^\prime}^< \tau) \hat{\sigma}_{+} + g^*(\Lambda_{\Omega^\prime}^< \tau) \hat{\sigma}_{-} \bigr) \Bigr\}
\end{aligned} 
\end{equation}
where $g(\Lambda_{\Omega^\prime}^<\tau) = 4 \Lambda_\Delta^{<}\sin^2(\sfrac{\Lambda_{\Omega^\prime}^<\tau}{2}) + 2i \Lambda_{\Omega^\prime}^< \sin(\Lambda_{\Omega^\prime}^< \tau)$ with generalized Rabi frequency $ \Lambda_{\Omega^\prime} = \sqrt{\Lambda_\Omega^2+\Lambda_\Delta^2} \equiv \lambda \Lambda_{\Omega^\prime}^< $ and $\Lambda_\Omega = \Omega/\omega_{eg} = 2\lambda $.
The second-order approximation for the effective dynamics is given by
\begin{equation}\label{eq:ex:Rabi_CR_Ueff_PT}
    \hat{U}_\mathrm{eff}^{[2]}(\tau;\lambda) \simeq \hat{U}_{\mathrm{eff},0}^\prime(\tau) \hat{U}_{\mathrm{eff}}^{\prime [1]}(\tau;\lambda) + \mathcal{O}(\lambda^2)
\end{equation}
where 
\begin{equation}
\hat{U}_{\mathrm{eff}}^{\prime [1]}(\tau;\lambda) \simeq \exp(-i \lambda \int^\tau \! \hat{H}_{I,\mathrm{eff}}^{\prime (2)}(\tau^\prime) \; d\tau^\prime)
\end{equation}
is the result from $\tau$-time exponential perturbation theory.

%% == PLOT == %%
\begin{figure}[!tp]
    \centering
    % \includegraphics[width=8.6cm]{plots/Raman_plot.pdf}
    %\resizebox{1.01\columnwidth}{!}{
    \includestandalone[width=\columnwidth]{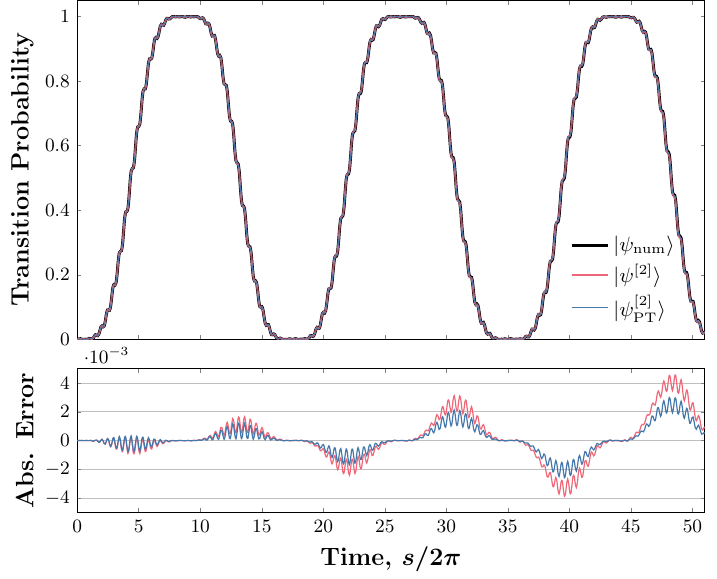}
    %}
    \caption{
        \textbf{Near-Resonant Rabi Oscillations with Counter-Rotating Terms.}  
        The qubit is initialized in the superposition state $\ket{\psi_0} = \frac{1}{\sqrt{2}}(\ket{g} + \ket{e})$ and weakly driven with amplitude $\lambda = 0.05$.  
        A near-resonant detuning $\Lambda_\Delta \simeq 1.14\lambda$ is chosen to eliminate the second-order light shift due to counter-rotating terms, thereby isolating higher-order contributions to the dynamics.  
        Both the non-perturbative (red) and perturbative (blue) second-order QAT dynamics closely match the full numerical solution, even beyond the formal regime of guaranteed validity.
    }
    \label{fig:Rabi_NearResonant}
\end{figure}

\subparagraph{Approximate QAT Dynamics.}
We complete the process by returning the unitary-preserving truncated QAT interaction propagator to second order, $\hat{U}_I ^{[2]}(s;\lambda) = \exp(-i \, \hat{\Phi}^{[1]}) \, \hat{U}_\mathrm{eff}^{[2]}(\tau;\lambda)|_{\tau=\lambda s}$,
where the dynamical phase contribution is given by \eqref{eq:ex:twotime_Rabi_QAT_dynphase_o1} and $\hat{U}_\mathrm{eff}^{[2]}(\tau;\lambda)$ is exactly and approximately given by \eqref{eq:ex:Rabi_CR_Ueff} and \eqref{eq:ex:Rabi_CR_Ueff_PT}, respectively.
Hence, for the system prepared in the state $\ket{\psi_0}$ at time $s=0$, the quantum state in the interaction picture evolves as $\ket{\psi_I^{[2]}(s,s_0=0)} ={} \hat{U}_I^{[2]}(s) \hat{U}_I^{\dagger [2]}(s_0) \ket{\psi_0} 
    \equiv{} \hat{U}_I^{[2]}(s,s_0) \ket{\psi_0}$.
The exact and the perturbation theory results for the truncated effective dynamics are compared against the numerical result in Figure~\eqref{fig:Rabi_NearResonant}.

\subsubsection*{Example 2: Near-Resonant Two-Photon Raman Transition}\label{sec:qat_multiplescales:Raman_twotime}

We revisit the two-photon Raman process described in section~\eqref{sec:qat:ex_Raman}, but now allowing the two-tone beat note to be slightly detuned from resonance.
As before, the interaction Hamiltonian is
\begin{equation}\label{eq:qat_multiplescales:Raman_interaction_Hamiltonian}
\hat{H}_I(s;\lambda) = \lambda \, \sum_{k=1,2} \left( \bar{\Omega}_k(s)  \ket{r}\bra{k} + h.c. \right).
\end{equation}
However, the normalized matrix elements
\begin{equation}\label{eq:qat_multiplescales:Raman_matrix_element}
    \bar{\Omega}_k(s) = 
\frac{\Omega_k }{2\Omega_{\mathrm{max}}}e^{i \phi_k} e^{-i \Lambda_{\Delta_k} s}  
\end{equation}
are assumed to have unequal detuning $\Lambda_{\Delta_k} = \Lambda_{\omega_k} - \Lambda_{\omega_0}^{(rk)}$ from the intermediate state $\ket{r}$.
As before, the intermediate state is driven far from resonance ($|\Lambda_{\Delta_k}| \gg \lambda$) yet the two-tone beat note $|\Lambda_{\Delta_1} - \Lambda_{\Delta_2}| = |(\Lambda_{\omega_1} - \Lambda_{\omega_2}) - \Lambda_{\omega_0}^{(21)}| \lesssim \lambda$ such that the difference is close to resonance with the transition frequency between states $\ket{1}$ and $\ket{2}$.
Using Table \eqref{table:twotime_freq} we parameterize the small relative detuning as $\Lambda_{\Delta_1}-\Lambda_{\Delta_2} = \lambda \Lambda_{\delta}$.
For convenience we use the symmetric parametrization $\Lambda_{\Delta_k} = \Lambda_\Delta \pm \lambda \Lambda_{\delta}/2$ for large $|\Lambda_\Delta| \gg \lambda > 0$.
The two-time interaction Hamiltonian is returned by using the two-time decomposition $\Lambda_{\Delta_k} s \equiv  \Lambda_\Delta \sigma \pm \Lambda_{\delta} \tau/2$ in \eqref{eq:qat_multiplescales:Raman_matrix_element}.

The leading non-zero contribution is the second-order QAT results
\begin{subequations}
\begin{align}
		\hat{\Phi}^{(1)}(\sigma,\tau;\lambda) &= \int^\sigma d\sigma' \; \hat{H}_I^{(1)}(\sigma',\tau) \\
  \hat{H}^{(2)}_\mathrm{eff}(\tau) &=  \hat{H}_\mathrm{R}(\tau) + \hat{H}_\mathrm{LS}  \label{eq:qat:Raman_effective_Hamiltonian}
\end{align}
\end{subequations}
where, using the shorthand notation $\bar{\Omega}_k \equiv \bar{\Omega}_k(0)$,
\begin{equation}
    \hat{H}_\mathrm{R}(\tau) = \frac{\bar\Omega_1 \bar\Omega_2^*}{4\Lambda_\Delta} e^{-i     \Lambda_{\delta} \tau} \ket{2}\bra{1} + h.c. ,
\end{equation}
which describes a slightly detuned stimulated Raman transition between states $\ket{1} \leftrightarrow \ket{2}$.
The second-order level shifts are given by $\hat{H}_\mathrm{LS} = \hat{H}_{\mathrm{LS},1} + \hat{H}_{\mathrm{LS},2}$ where
\begin{equation}
\begin{aligned}
        \hat{H}_{\mathrm{LS},1} ={}& \frac{1}{\Lambda_\Delta} \frac{|\bar\Omega_1|^2}{4}  \left( \ket{1} \bra{1} - \ket{r}\bra{r} \right) \\
        \hat{H}_{\mathrm{LS},2} ={}&  \frac{1}{\Lambda_\Delta} \frac{|\bar\Omega_2|^2}{4}  \left(\ket{2} \bra{2} - \ket{r}\bra{r}\right) .
\end{aligned}
\end{equation}
The third-order contributions only introduce adiabatic corrections given by
\begin{subequations}
    \begin{align}
        \hat{\Phi}^{(2)}(\sigma,\tau) ={}& -\int d\sigma \; \partial_\tau \hat{\Phi}^{(1)}(\sigma,\tau) \\
            \hat{H}_\mathrm{eff}^{(3)}(\tau) ={}& -\frac{1}{2} \timeavg{[-i \hat{\Phi}^{(2)}, \hat{H}_I^{(1)}]}{\sigma} =\hat{H}_\mathrm{LS}^{\mathrm{ad}}
    \end{align}
\end{subequations}
where
\begin{subequations}
    \begin{align}
    & \begin{aligned}
        \hat{H}_\mathrm{LS}^\mathrm{ad} ={}& -\frac{\Lambda_\delta}{2\Lambda_\Delta} \left[  \hat{H}_{\mathrm{LS},1} -  \hat{H}_{\mathrm{LS},2} \right].
    \end{aligned}
    \end{align}
\end{subequations}
The third-order effective dynamics are exactly solvable yielding
\begin{equation}\label{eq:qat:Raman_eff_propagator}
	\hat{U}_\mathrm{eff}^{[3]}(\tau;\lambda) = \exp(i \Lambda_{\delta} \hat{S}_{\mathrm{Z}_2} \tau) \exp(-i \hat{H}_\mathrm{R}^\prime \tau)
\end{equation}
where
\begin{equation}
    \hat{H}_\mathrm{R}^\prime = \Lambda_{\delta}\hat{S}_{\mathrm{Z}_2} + \lambda [\hat{H}_\mathrm{LS}+\hat{H}_\mathrm{R}(0)] + \lambda^2 \hat{H}_\mathrm{LS}^\mathrm{ad}
\end{equation}
where $\hat{S}_{\mathrm{Z}_2} = \frac{1}{2}(\ketbra{1}{1} - \ketbra{2}{2})$.
Note that the freedom to choose a symmetric frequency decomposition is an example of the non-uniqueness of the two-timescale expansion.
Alternatively, the parametrization $\Lambda_{\Delta_1} = - \Lambda_\Delta + \lambda \Lambda_{\delta}$ and $\Lambda_{\Delta_2} = - \Lambda_\Delta$ yields a different expansion that is equally valid to same order of approximation.
This discrepancy will be addressed in the following section.

\subsection{Partitioned Expansion by Timescale Separation (PETS) Framework}\label{sec:qat_multiplescales:PETS}

When the interaction Hamiltonian involves a large number of modes, applying frequency decomposition becomes increasingly cumbersome due to the complexity of tracking individual contributions.
In the companion paper \cite{barajasMultiTimescaleCoherentControl2025}, we introduced a simple, inductive method for constructing a two-timescale QAT framework using the Partitioned Expansion by Timescale Separation (PETS) approach.
The PETS framework leverages the RWA as a familiar regularization technique to account for resonant and low-frequency effects, enabling effective descriptions that extend beyond first-order approximations.
Mathematically, the PETS approach is entirely consistent with the two-timescale derivative expansion technique without requiring explicit determination of regularization parameters, provided $\lambda \le \Lambda_\mathrm{c} < 1$.
Although less mechanical than the two-timescale derivative expansion, PETS offers a more intuitive and practical alternative for handling multi-timescale dynamics in complex systems.

We begin from the QAT homological equation given by eq.~\eqref{eq:qat:QAT_homological}.
Consider that for the most general almost-periodic interaction Hamiltonian in \eqref{eq:qat:Fourier_basis} that the auxiliary operator will also be almost-periodic with the same base frequencies $\Lambda_k \in \vec{\Lambda}$.
Without loss of generality, we separate the $n$th-order auxiliary operator in terms of resonant (with subscript $0$), fast ($>$), and slow ($<$) Fourier modes:
\begin{equation}\label{eq:qat_multiplescales:aux_Fourier_basis}
	\hat{\mathcal{H}}_\Phi ^{(n)}(s) = \hat{\mathcal{H}}_{\Phi,0}^{(n)} + \hat{\mathcal{H}}_{\Phi,>} ^{(n)}(s) +\hat{\mathcal{H}}_{\Phi,<} ^{(n)}(s) ,
\end{equation}
where
\begin{equation}\label{eq:qat_multiplescales:aux_Fourier_terms}
        \hat{\mathcal{H}}_{\Phi,\gtrless}^{(n)}(s) ={} \sum_{\mathclap{\vec{\alpha}_k \in F_n^\gtrless}} e^{-i \vec{\alpha}_k \cdot \vec{\Lambda} s} \, \hat{\mathcal{H}} ^{(n)}_{\Phi,k} + h.c.
\end{equation}
are the time-dependent contributions and $\hat{\mathcal{H}}_{\Phi,0}^{(n)}$ collects the resonant, time-independent terms.
We introduced the set $F_n$ of unique integer vectors $\vec{\alpha}_k \in \mathbb{Z}^q$ that generate the base as well as the (positive) sum and difference frequencies appearing at $n$th order.
The superscript ``$\gtrless$'' on the set denotes whether the integer vectors $\vec{\alpha}_k$ generate a fast ($\vec{\alpha}_k \cdot \vec{\Lambda} > \lambda$) or slow ($\vec{\alpha}_k \cdot \vec{\Lambda} \le \lambda$) frequency.

As before, we seek to maintain a non-secular theory. 
And, in addition, we desire the expansion to be asymptotically valid to all orders, even when $F_n^< \neq 0$.
To this end, we define the RWA as the low-pass filter procedure
\begin{equation}\label{eq:qat_multiplescales:rwa_lowpass_filter}
    \timeavg{\hat{A}(s)}{\mathrm{rwa}} = \int_{-\infty}^\infty ds' f(s-s') \hat{A}(s')
\end{equation}
for an idealized brick-wall response $f(s) = \frac{\Lambda_\mathrm{c}}{2\pi} \mathrm{sinc}(\Lambda_\mathrm{c} s)$ with a high-frequency cutoff $\Lambda_c = \lambda$.
In the PETS approach, the homological equation is regularized by separating the fast from the slow timescale effects by requiring
\begin{equation}
    \timeavg{\hat{\Phi}^{(n)}(s)}{\mathrm{rwa}} = 0,
\end{equation}
which is satisfied by requiring
\begin{equation}\label{eq:qat_multiplescales:Heff_PETS}
\begin{aligned}
    \hat{H}_{I,\mathrm{eff}}^{(n)}(s) ={}& \timeavg{\hat{\mathcal{H}}_\Phi ^{(n)}(s)}{\mathrm{rwa}} =  \hat{\mathcal{H}}_{\Phi,0}^{(n)} +\hat{\mathcal{H}}_{\Phi,<} ^{(n)}(s)
\end{aligned}
\end{equation}
guaranteeing an asymptotically valid expansion.
If we define $\vec{\Lambda}_\mathrm{eff}>0$ to be the frequency vector of $\hat{H}_{I,\mathrm{eff}}(s;\lambda)$, then each component $\Lambda_k \in \vec{\Lambda}_\mathrm{eff}$ satisfies $0 < \Lambda_k \le \lambda$.
Hence, the slow timescale $\tau=\lambda s$ from the two-timescale expansion in section \eqref{sec:qat_multiplescales:twotime} also naturally arises in the PETS approach.

We observe that the concept of the RWA as a low-pass filter appears organically from the requirement of asymptotic validity. 
A similar approach was proposed by James \textit{et. al} \cite{jamesEffectiveHamiltonianTheory2007} to generate an effective Hamiltonian from a time-ordered Dyson series.
However, the approach suffered from non-unitary artifacts that do not appear in a Magnus-type expansion.
Moreover, the expansion is kept exact by generating the high-frequency fast propagator from the regularized homological equation
\begin{equation}\label{eq:qat_multiplescales:PETS_homological_eqn}
\begin{aligned}
    \frac{d}{ds}\hat{\Phi}^{(n)}(s) ={}& \hat{\mathcal{H}}_{\Phi}^{(n)}(s) - \hat{H}_{I,\mathrm{eff}}^{(n)}(s) =  \hat{\mathcal{H}}_{\Phi,>} ^{(n)}(s)
\end{aligned}
\end{equation}
with the theory proceeding otherwise the same as in Sec.~\eqref{sec:qat}.
We remark that while we initially assumed an RWA cutoff frequency $\Lambda_c =\lambda$, comparing with the two-timescale derivative approach reveals this to be the minimal bound of the broader constraint $\lambda \le \Lambda_\mathrm{c} <1$, which introduces a clear separation between fast and slow dynamics.

% We have demonstrated that a renormalized Magnus expansion with a PETS approach provides a general unitary framework for treating multi-timescale driven quantum systems.
% Additionally, the RWA-based regularization implies the presence of at least two timescales: fast dynamics on the order of $s \lesssim \mathcal{O}(1)$ and slower dynamics on the order of $s \gtrsim \mathcal{O}(1/\lambda)$, determined by the natural timescale of the perturbation parameter $\lambda$. 
% Leveraging the algorithmic Magnus expansion, the PETS approach enables arbitrary-order precision in capturing system dynamics.
% In this framework, the RWA serves as a familiar regularization technique that accounts for resonant and low-frequency effects at higher orders, extending beyond simple coarse-grained descriptions. 

It is important to note, however, that the effective Hamiltonian may not have a uniquely defined form: the RWA low-pass filter relies on a system-specific frequency cutoff and the mode spectrum depends on the physical parameters of the problem. 
This constraint poses minimal challenges in constructing an appropriate effective system, provided the characteristic timescales of the interaction Hamiltonian are well understood. 
% As we will now demonstrate, from a mathematical point of view, the PETS approach is consistent with the well-studied two-timescale derivative expansion technique, provided $\lambda \le \Lambda_\mathrm{c} < 1$.

% To rigorously justify the PETS procedure, we will now demonstrate that it is entirely equivalent to the two-timescale derivative expansion approach to the same order of approximation, which is the main result of this section.
% More specifically, we will now prove that the adiabatic corrections in \eqref{eq:qat_multiplescales:QAT_homological} correspond to the coefficients of the series expansion of the mode amplitudes of $\hat{\Phi}^{(n)}(s)$ calculated with the PETS method, expanded in powers of $\lambda$.
To rigorously justify the PETS procedure, we establish its equivalence to the two-timescale derivative expansion at a given order of approximation.
This equivalence represents the central result of this section.
Specifically, we prove that the adiabatic corrections in \eqref{eq:qat_multiplescales:QAT_homological} align with the coefficients of the series expansion for the mode amplitudes of $\hat{\Phi}^{(n)}(s)$, calculated using the PETS approach.
This relationship between the adiabatic corrections and the series coefficients in powers of $\lambda$ establishes the consistency of the two approaches.
From the homological equation \eqref{eq:qat_multiplescales:PETS_homological_eqn} for $n=1$, solving for $\hat{\Phi}^{(1)}(s)$ and expanding the mode amplitudes in powers of $\Lambda_c = \lambda \Lambda_c^<$ yields
\begin{equation}
    \begin{aligned}
        \hat{\Phi}^{(1)}(s) ={}& \sum_{\Lambda_k > \Lambda_\mathrm{c}} \frac{1}{i \Lambda_k} e^{i \Lambda_k s} \hat{h}_k + h.c. \\
        ={}& \sum_{\Lambda_k > \Lambda_\mathrm{c}} \frac{1}{i \Lambda_{k}^>} (1- \Lambda_c \frac{\Lambda_{\omega}^<}{\Lambda_{\omega}^>} + \ldots)e^{i \Lambda_k s} \hat{h}_k + h.c.
    \end{aligned}
\end{equation}
where we've inserted the frequency decomposition \eqref{eq:qat_multiplescales:freq_decomposition} from the two-timescale derivative approach.
On the other hand, adiabatic passage of $\tau$-time in the two-timescale derivative expansion implies
\begin{equation}
    \begin{aligned}
        \hat{\Phi}^{(1)}(\sigma,\tau) ={}& \sum_{\Lambda_k > \Lambda_\mathrm{c}}\frac{1}{i \Lambda_{k}^>} e^{i (\Lambda_k^> \sigma + \Lambda_k^< \tau)} \hat{h}_k + h.c.,
        % \hat{\Phi}^{(2)}_\mathrm{ad}(\sigma,\tau)={}& - \Lambda_c^< \int^\sigma d\sigma^\prime \; \partial_\tau \hat{\Phi}^{(1)}(\sigma^\prime,\tau) \\
        % ={}& - \Lambda_c^< \left( \frac{\Lambda_{\omega}^<}{i (\Lambda_{\omega}^>)^2} e^{i \Lambda_\omega s} \hat{h} + h.c. \right)
    \end{aligned}
\end{equation}
% where $\hat{\Phi}^{(2)}_\mathrm{ad}(\sigma,\tau)$ is the adiabatic correction term at next order.
which by comparison reveals $\hat{\Phi}^{(1)}(s) = \hat{\Phi}^{(1)}(\sigma,\tau) + \mathcal{O}(\lambda^2)$ such that both are valid descriptions to same order of approximation.
Hence, if we define the adiabatic correction operator $D_\mathrm{ad}:\hat{A}(\sigma,\tau) \mapsto -\Lambda_c^< \int^\sigma d\sigma^\prime \partial_\tau \hat{A}(\sigma^\prime,\tau)$, then one can easily show that
\begin{equation}
\hat{\Phi}^{(1)}(s) = \sum_{k =0}^\infty \lambda^k D_\mathrm{ad}^k \circ \hat{\Phi}^{(1)}(\sigma,\tau)|_{\sigma=s,\tau=\Lambda_c s},
\end{equation}
which coincides with the result above.
Extending to any $n >1$ follows trivially by induction.
Hence, for $\lambda \le \Lambda_\mathrm{c} <1$, the results from the PETS approach (left) are related to the two-timescale derivative expansion (right) as follows:
\begin{equation}
    \begin{aligned}
        \hat{H}_\mathrm{eff}^{(1)}(s) ={}& \hat{H}_\mathrm{eff}^{(1)}(\tau) \\
        \hat{H}_\mathrm{eff}^{(n)}(s) ={}& \hat{H}_\mathrm{eff}^{(n)}(\tau) + \mathcal{O}(\lambda), \quad n\ge 2 \\
        \hat{\Phi}^{(n)}(s) = {}& \hat{\Phi}^{(n)}(\sigma,\tau) +\mathcal{O}(\lambda), \quad n\ge 1
    \end{aligned}
\end{equation}
such that the two methods are equivalent to the same order of approximation.
It follows that the multi-timescale averaging procedures are related by
\begin{equation}
    \timeavg{\hat{A}(s)}{\mathrm{rwa}} = \timeavg{\hat{A}(\sigma,\tau)}{\sigma} + \mathcal{O}(\lambda),
\end{equation}
where the RWA filters out fast Fourier modes with implicit dependence on the fast-time $\sigma$.

The difference between the methods can be summarized as follows.
The PETS approach is an inductive method, requiring the homological equation to be assessed and regularized iteratively, order-by-order.
In contrast, the derivative expansion is best suited as a deductive method that enforces validity prior to any calculations by determining a suitable frequency decomposition. We note that the frequency decomposition in a derivative expansion approach can also be performed inductively in a process requiring additional steps compared to the PETS approach.
Nonetheless, it may be advantageous in scenarios where one seeks to engineer specific interactions by introducing arbitrary drives, such as through continuous-time pulse shaping techniques \cite{ribeiroSystematicMagnusBasedApproach2017,sametiStrongcouplingQuantumLogic2021}.

\subsubsection*{Example 1: Near-Resonant Rabi Problem with PETS Approach}
We revisit the near-resonant Rabi problem explored in \eqref{sec:qat_multiplescales:Rabi_twotime}, this time applying the PETS approach to study the long-time dynamics.
We reiterate, in the PETS approach an explicit frequency decomposition is not required.
Instead, a timescale separation will be addressed by effectively applying the RWA at each order while still maintaining an exact expansion.

\subparagraph{Interaction Picture.} As before we have the interaction picture Hamiltonian
\begin{equation}
	\hat{H}_I^{(1)}(s) = \left(e^{-i \Lambda_\Delta s} + e^{i \Lambda_\Sigma s} \right) \hat{\sigma}_+ + h.c.
\end{equation}
where we assume the detuning parameter to be nearly resonant (i.e. $\Lambda_\Delta \le \Lambda_\mathrm{c}$) such that $\Lambda_\Sigma \approx 2 \gg \Lambda_\mathrm{c}$.

\subparagraph{Regularized QAT Homological Equation.} Using Table~\ref{table:qat:auxiliary_hamiltonian} and eq.~\eqref{eq:qat_multiplescales:QAT_time_average}, for $n=1$ we have
\begin{equation}
            \hat{\mathcal{H}}_{\hat{\Phi}}^{(1)}(s) ={} \hat{H}_I^{(1)}(s) = \hat{H}_{I,>}^{(1)}(s) + \hat{H}_{I,<}^{(1)}(s)
\end{equation}
where
\begin{equation}
\begin{aligned}
        \hat{\mathcal{H}}_{\hat{\Phi},>}^{(1)}(s) \equiv{}& \hat{H}_{I,>}^{(1)}(s) = e^{i \Lambda_\Sigma s} \, \hat{\sigma}_+ + h.c.\\
        \hat{\mathcal{H}}_{\hat{\Phi},<}^{(1)}(s) \equiv{}& \hat{H}_{I,<}^{(1)}(s) = e^{-i \Lambda_\Delta s} \, \hat{\sigma}_+ + h.c.,
\end{aligned}
\end{equation}
separated with respect to fast and slow modes, respectively.
Applying a low-pass filter (i.e. RWA), the first-order contributions are given by
\begin{subequations}
	\begin{align}
        \hat{H}_{I,\mathrm{eff}}^{(1)}(s) ={}& \timeavg{\hat{\mathcal{H}}_{\Phi}^{(1)}(s)}{\mathrm{rwa}} = e^{-i \Lambda_\Delta s} \, \hat{\sigma}_+ + h.c. \\
        \begin{split} 
		\hat{\Phi}^{(1)}(s) ={}&  \int^s \! d{s'} \: {\hat{\mathcal{H}}_{\hat{\Phi},>}^{(1)}(s')} = \frac{e^{i \Lambda_\Sigma s}}{i \Lambda_\Sigma}  \, \hat{\sigma}_+ + h.c.
        \end{split}
	\end{align}
\end{subequations}
% where we've applied the RWA low-pass filter.
Repeating the algorithm to next two orders we have
\begin{subequations}
	\begin{align}
        \hat{H}_{I,\mathrm{eff}}^{(2)} ={}&\frac{1}{\Lambda_\Sigma} \, \hat{\sigma}_z \\
        \begin{split}
	\hat{\Phi}^{(2)}(s) ={}& \frac{2}{\Lambda_\Sigma (\Lambda_\Delta+\Lambda_\Sigma) } \sin((\Lambda_\Delta + \Lambda_\Sigma) s) \, \hat{\sigma}_z.
        \end{split}
	\end{align}
\end{subequations}
and
\begin{subequations}
	\begin{align}
        \hat{H}_{I,\mathrm{eff}}^{(3)}(s) ={}& -\frac{1}{\Lambda_\Sigma(\Lambda_\Delta+\Lambda_\Sigma)} e^{-i \Lambda_\Delta s} \, \hat{\sigma}_+ + h.c. \\
        \begin{split}
	\hat{\Phi}^{(3)}(s) ={}& \frac{1}{6 \Lambda_\Sigma^3} \Bigl[ -i \, \hat{\sigma}_{+} \Bigl(\frac{12 \Lambda_\Sigma ^2 e^{-i s (2\Lambda_\Delta +\Lambda_\Sigma )}}{(\Lambda_\Delta +\Lambda_\Sigma ) (2 \Lambda_\Delta +\Lambda_\Sigma )}   \\
    &{} + \frac{6 \Lambda_\Sigma ^2 e^{i s (\Lambda_\Delta +2\Lambda_\Sigma )}}{(\Lambda_\Delta +\Lambda_\Sigma ) (\Lambda_\Delta +2 \Lambda_\Sigma )} \\
    &{} -\Bigl(8-\frac{12 \Lambda_\Sigma }{\Lambda_\Delta +\Lambda_\Sigma}\Bigr) e^{i s \Lambda_\Sigma}  \Bigr) + h.c. \Bigr].
        \end{split}
	\end{align}
\end{subequations}
For this example we find that the second-order PETS results match identically with the results from the two-timescale derivative expansion approach.
At higher-orders, performing a series expansion in powers of $\lambda$ of the mode amplitudes in the results above  using $\Lambda_\Delta = \lambda \Lambda_\Delta^<$ shows full agreement with the two-timescale derivative expansion results to same order of approximation as expected.
The final two steps shown in section \eqref{sec:qat_multiplescales:Rabi_twotime} are omitted as the remaining process is the same as before.

\subsubsection*{Example 2: Near-Resonant Two-Photon Raman Transition with PETS Approach}
We return to the near-resonant two-photon Raman process described in section~\eqref{sec:qat_multiplescales:Raman_twotime}, now applying the PETS approach.
% The detuning $\Lambda_{\Delta_k} = \Lambda_{\omega_k} - \Lambda_{\omega_0}^{(rk)}$ and far off-resonant component $\Lambda_{\Sigma_k} = \Lambda_{\omega_k} + \Lambda_{\omega_0}^{(rk)} \gg \Lambda_{\Delta_k}$ are set by the dimensionless drive $\Lambda_{\omega_k}$ and the energy level gap $\Lambda_{\omega_0}^{(rk)} > 0$ between states $\ket{k} \leftrightarrow \ket{r}$ for $k=1,2$. 
% In the far-detuned regime we may safely ignore the counter-rotating terms to first-order approximation as leading order corrections are proportional to $\lambda/\Sigma_k \lll \lambda/\Delta_k$.
Using the same conditions $|\Lambda_{\Delta_k}| \gg \lambda$ and $|\Lambda_{\Delta_1} - \Lambda_{\Delta_2}| = |(\Lambda_{\omega_1} - \Lambda_{\omega_2}) - \Lambda_{\omega_0}^{(21)}| \lesssim \lambda$, the leading non-zero contribution are the second-order QAT results
\begin{subequations}
\begin{align}
		\hat{\Phi}^{(1)}(s;\lambda) &= \int^s ds' \; \hat{H}_I^{(1)}(s') \\
  \hat{H}^{(2)}_\mathrm{eff}(s) &= \timeavg{\hat{\mathcal{H}}_{\Phi}^{(2)}(s)}{\mathrm{rwa}} = \hat{H}_\mathrm{R}(s) + \hat{H}_\mathrm{LS}  \label{eq:qat_multiplescales:Raman_PETS_effective_Hamiltonian}
\end{align}
\end{subequations}
where, using the shorthand notation $\bar{\Omega}_k \equiv \bar{\Omega}_k(0)$,
\begin{equation}
    \hat{H}_\mathrm{R}(s) = \frac{\bar\Omega_1 \bar\Omega_2^*\Lambda_{\Delta_1 + \Delta_2}}{8\Lambda_{\Delta_1}\Lambda_{\Delta_2}} e^{-is     \Lambda_{\Delta_1-\Delta_2}} \ket{2}\bra{1} + h.c.,
\end{equation}
is the slightly-detuned Raman transition while the level shifts $\hat{H}_\mathrm{LS} = \hat{H}_{\mathrm{LS},1} + \hat{H}_{\mathrm{LS},2}$ are
\begin{equation}
\begin{aligned}
        \hat{H}_{\mathrm{LS},1} ={}& \frac{1}{\Lambda_{\Delta_1}} \frac{|\bar\Omega_1|^2}{4}  \left( \ket{1} \bra{1} - \ket{r}\bra{r} \right) \\
        \hat{H}_{\mathrm{LS},2} ={}&  \frac{1}{\Lambda_{\Delta_2}} \frac{|\bar\Omega_2|^2}{4} \left(\ket{2} \bra{2} - \ket{r}\bra{r}\right).
\end{aligned}
\end{equation}
The second-order effective dynamics are exactly solvable yielding $\hat{U}_\mathrm{eff}^{[2]}(\tau;\lambda) = \exp(i \hat{H}_\mathrm{D} s) \exp(-i[\hat{H}_\mathrm{R} + \hat{H}_\mathrm{eff}^{[2]}(0)] s)$.

To higher orders, we find that all odd order effective Hamiltonian contributions $\hat{H}_\mathrm{eff}^{(2k+1)} = 0$ for integer $k$, vanishing as expected of a two-photon process.
Here, all off-resonant single-photon processes to the intermediate state $\ket{r}$ are accounted for in the dynamical phase operator.
Expanding $\lambda^2 \hat{H}_\mathrm{eff}^{(2)}(s)$ in powers of $\lambda$ with $\Lambda_{\Delta_k} = \Lambda_\Delta \pm \lambda \Lambda_\delta/2$ to third-order yields the results from the multi-timescale derivative expansion, including the third-order adiabatic correction.

%% == PLOT == %%
\begin{figure}[!tp]
    \centering
    % \includegraphics[width=8.6cm]{plots/Raman_plot.pdf}
    %\resizebox{1.01\columnwidth}{!}{
    \includestandalone[width=\columnwidth]{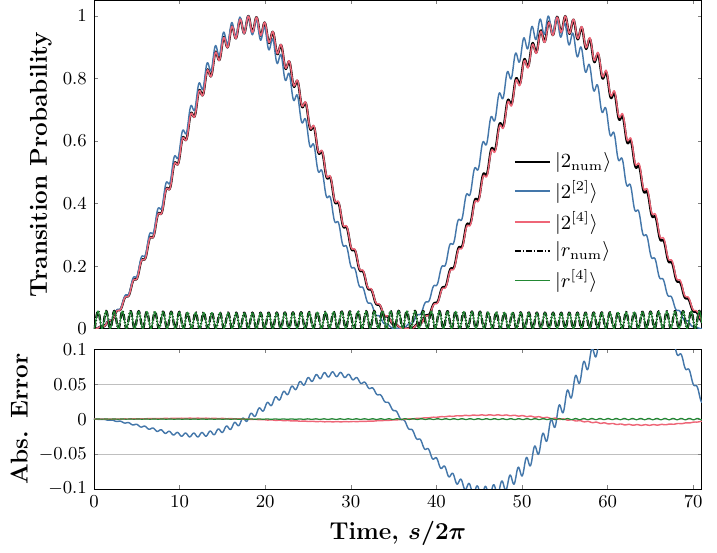}
    %}
    \caption{
    \textbf{Near-Resonant Raman Transition.}  
    The system is initialized in the ground state $\ket{\psi_0} = \ket{1}$.  
    The pump and probe beams are driven with moderate coupling strengths $\Omega_1 = \Delta_1 / 4$ and $\Omega_2 = \Delta_1 / 5$, respectively, with detuning from the intermediate state $\ket{r}$ set by $\Delta_1 = 2\pi \times \SI{10}{MHz}$.  
    The frequency difference $\Delta = \Delta_1 - \Delta_2$ is chosen to cancel second-order differential light shifts, ensuring $\bra{1}\hat{H}_\mathrm{eff}^{[2]}\ket{1} = \bra{2} \hat{H}_\mathrm{eff}^{[2]} \ket{2}$.  
    Second-order QAT results (blue) agree with the numerical solution over half a Rabi cycle but deviate at longer times.  
    In contrast, the next non-zero correction (red) shows rapid convergence and remains accurate over multiple cycles.
    }
    \label{fig:Raman_NearResonant}
\end{figure}

As in the resonant case, the QAT results for the slightly-detuned Raman process using the PETS approach coincide with a closed-form solution.
The exact dynamical phase operator can be found using the ansatz
\begin{equation} 
\hat{\Phi}(s;\lambda) = \lambda \, \sum_{k=1,2} \theta_k \left( \int^s ds' \,\bar{\Omega}_k(s) \right) \ket{r}\bra{k} + h.c.
\end{equation}
where $\theta_k = \alpha_k \sfrac{\sqrt{\Lambda_{\Delta_1}\Lambda_{\Delta_2}} } {\Omega^\prime_{\mathrm{rms}}} \arctan(\Omega^\prime_{\mathrm{rms}}/\sqrt{\Lambda_{\Delta_1}\Lambda_{\Delta_2}})$ and $\Omega^\prime_{\mathrm{rms}} = \sqrt{\alpha_1^2 \Lambda_{\Delta_2}^2 |\Omega_1|^2 + \alpha_2^2 \Lambda_{\Delta_1}^2|\Omega_2|^2}/\sqrt{\Lambda_{\Delta_1}\Lambda_{\Delta_2}}$. 
The parameters $\alpha_k = 1 \pm(\Lambda_{\Delta_1} - \Lambda_{\Delta_2})\alpha_k^\prime$ are real-valued and are unity for the two-photon resonant case $\Lambda_{\Delta_1} = \Lambda_{\Delta_2}$, yielding the resonant QAT solution.
The primed parameters $\alpha^\prime_k$ satisfy a cubic equation generated by requiring $\bra{1} \hat{H}_\mathrm{eff} \ket{r} =\bra{2} \hat{H}_\mathrm{eff} \ket{r} =0$ and are equivalent under exchange of indices (i.e $\alpha^\prime_2 \equiv \alpha^\prime_1(\Lambda_{\Delta_1} \leftrightarrow\Lambda_{\Delta_2}$, $\bar\Omega_1 \leftrightarrow \bar\Omega_2)$).
The resulting effective Hamiltonian $\hat{H}_\mathrm{eff}(s;\lambda) = \lambda^2 [\sum_{k=1,2} h_k(\lambda) \hat{H}_{\mathrm{LS},k} + (\bar\Omega_\mathrm{R}(\lambda) e^{-is \Lambda_{\Delta_1-\Delta_2}} \ketbra{2}{1} +h.c.)]$ is block diagonal and traceless and reduces to the two-photon resonant solution for $\Lambda_{\Delta_1}=\Lambda_{\Delta_2}$.
% We take the real root valid for all parameter regimes yielding
% \begin{equation}
%     \alpha^\prime_1 = 2\sqrt{\frac{-p}{3}}\cos \left(\frac{1}{3}\arccos\left(\frac{3q}{2p}\sqrt{\frac{-3}{p}}\right) \right)- \frac{b}{3a}
% \end{equation}
% for the depressed cubic parameters $p = (3ac-b^2)/3a^2$ and $q = (2b^3-9abc+27a^2d)/27a^3$ where $d = 1$ and
% \begin{widetext}
% \begin{equation}
% \begin{aligned}
%     a = {}&\Lambda_{\Delta_1-\Delta_2}|\bar\Omega_2|^2(-16 \Lambda_{\Delta_1}\Lambda_{\Delta_2}^2|\bar{\Omega}_2|^2\Lambda_{\Delta_1-\Delta_2} + [\Lambda_{\Delta_1}|\bar{\Omega}_2|^2+ \Lambda_{\Delta_2}|\bar{\Omega}_1|^2 + 4\Lambda_{\Delta_1}\Lambda_{\Delta_2}\Lambda_{\Delta_1-\Delta_2}]^2) \\
%     b = {}& \Lambda_{\Delta_1-\Delta_2}(16 \Lambda_{\Delta_1}^2 \Lambda_{\Delta_2}|\bar\Omega_1|^2-\Lambda_{\Delta_1-\Delta_2}|\bar\Omega_2|^2)+[\Lambda_{\Delta_1}|\bar\Omega_1|^2+\Lambda_{\Delta_2}|\bar\Omega_2|^2-\Lambda_{\Delta_1-\Delta_2}(4\Lambda_{\Delta_1}\Lambda_{\Delta_2}+|\bar\Omega_2|^2)]^2 \\
%     c = {}&- 2(\Lambda_{\Delta_1}|\bar\Omega_1|^2+\Lambda_{\Delta_2}|\bar\Omega_2|^2) - 2 \Lambda_{\Delta_1}[\Lambda_{\Delta_1-\Delta_2}^2+\Lambda_{\Delta_2}^2]+\Lambda_{\Delta_1-\Delta_2}|\bar\Omega_2|^2.
% \end{aligned}
% \end{equation}
% \end{widetext}

% (see Appendix~\eqref{appendix:suN_Lie_algebra} for  $\mathfrak{su}(N)$ Lie algebra).
%The approximate total system dynamics given by \eqref{eq:qat:second_order_approx_dynamics} are compared with numerical results in Fig. \eqref{fig:raman}.

\subsection{Iterative Timescale Separation for Multi-Timescale Quantum Dynamics}\label{sec:qat_multiplescales:multitime}

In practical application, a two-timescale QAT is often sufficient to capture the system dynamics.
Nonetheless, it is straightforward to extend the two-timescale procedures, both PETS and the derivative expansion, to study multi-timescale dynamics.
A generalized multi-timescale procedure will allow us to to study a wideer range of time-dependent perturbations generating dynamics on increasingly slower timescales.
Due to the equivalence, throughout the following we will interchangeably refer to the PETS and the derivative expansion approach, depending on the aspects we wish to emphasize.

The following primarily serves to illustrate how slower timescale effects are, in fact, hidden in a two-timescale expansion.
Suppose we have $\hat{H}_I(s) \equiv \hat{H}_I(\sigma,\bm{\tau})$ generating dynamics on multiple different timescales with the system time %s(\sigma, \bm{\tau})$ 
parameterized by a fast time $\sigma$ and additional slow timescales $\bm{\tau} = \{\tau_j\}$ where $\tau_j = \Lambda_c^j \, s$ for integer $j = 1, \ldots, M$.
For simplicity, we will take the high-frequency cutoff $\Lambda_\mathrm{c} = \lambda$.
The additional slow timescales are already accounted for in the Fourier modes through the low-frequency component $\Lambda_k^< = \sum_{n\ge0} \lambda^n \Lambda_k^{(n)} \le 1$ defined in eq.~\eqref{eq:qat_multiplescales:freq_decomposition}.
The generalized phase in eq.~\eqref{eq:qat_multiplescales:Fourier_phase} expressed in terms of the set of slow-times $\{\tau_j\}$ becomes
\begin{equation}\label{eq:qat:Fourier_phase_multi}
\begin{aligned}
    \Lambda_{\omega_k} s ={}& (\Lambda_{\omega_k}^{>} + \lambda \Lambda^{(1)}_{\omega_k} + \lambda^2 \Lambda^{(2)}_{\omega_k} + \cdots ) s \\
    ={}& \Lambda_{\omega_k}^{>} \,\sigma + \underbracket{\Lambda^{(1)}_{\omega_k} \, \tau_1}_{\mathcal{O}(\lambda)} + \underbracket{\Lambda^{(2)}_{\omega_k} \, \tau_2}_{\mathcal{O}(\lambda^2)} + \cdots
\end{aligned}
\end{equation}
Hence, the generalization to a multi-timescale derivative expansion is quite straightforward (see appendix~\eqref{appendix:multi_timescale_expansion}).
% As in the two-timescale description, the slower timescale dynamics may be generated by near-resonant multi-modal interactions from higher-order processes.
As we will see, by using a Multiple PETS renormalization procedure, the additional timescales and frequency regularization parameters need not be explicitly defined.
However, their explicit reference will serve as a useful visual guide throughout the following.
% tool in the development of an iterative, Multiple PETS procedure for multi-timescale quantum systems.

Following the discussion above, applying the PETS procedure of Sec.~\eqref{sec:qat_multiplescales:PETS} to the (multi-timescale) interaction Hamiltonian yields the effective Hamiltonian $\hat{H}_{I,\mathrm{eff}}(\tau_1;\lambda) \equiv \hat{H}_{I,\mathrm{eff}}(\bm{\tau};\lambda)$ with $\tau_1$-time slow modes characterized by the rescaled frequency vector $\vec{\Lambda}_\mathrm{eff}/\lambda$ with components $0 < \Lambda_k^< \le 1$.
From \eqref{eq:qat:QAT_eff_system_slowtime}, we may express the effective equation as the ($\bm{\tau}$-time) perturbation problem
\begin{equation}\label{eq:qat_multiplescales:multitime_perturbation_problem}
\begin{split}
    i \! \left(\frac{1}{\lambda} \frac{d}{ds}\right)\hat{U}_{\mathrm{eff}}(\bm{\tau};\lambda) ={}& \Bigl( \hat{H}_{I,\mathrm{eff}}^{(1)}(\bm{\tau}) \; + \\
    & \! \sum_{n=2}^{\infty} \lambda^{n-1} \hat{H}_{I,\mathrm{eff}}^{(n)}(\bm{\tau}) \Bigr) \hat{U}_{\mathrm{eff}}(\bm{\tau};\lambda)
\end{split}
\end{equation}
where, unlike in a multi-timescale derivative expansion, the rescaled time-derivative $\lambda^{-1} d/ds \equiv d/d\tau_1$ ``hides'' the dependence on slower timescales.
%$\partial_{\tau_1} + \sum_{j=2}^{M} \lambda^{j-1} \partial_{\tau_j}$ (dropping $\partial_\sigma$ since $\hat{U}_I(\bm{\tau};\lambda)$ doesn't depend on $\sigma$). 
As a reasonable assumption, suppose the $\boldsymbol{\tau}$-time perturbation problem is not explicitly solvable.
As in Sec.~\eqref{sec:perturbation_problem}, if the ``unperturbed'' $\hat{H}_{I,\mathrm{eff}}^{(1)}(\bm{\tau})$ is solvable, then we may transform into an interaction picture depending only on the perturbation (see appendix~\eqref{appendix:effective_equation_PT} for details).
Let $\hat{U}_{0}^{\prime}(\bm{\tau})$ be the exactly solvable, ``unperturbed'' propagator solution to
\begin{equation}\label{eq:qat_multiplescales:multitime_unperturbed_problem_solution}
i \! \left(\frac{1}{\lambda} \frac{d}{ds}\right)\hat{U}_{0}^\prime(\bm{\tau}) ={} \hat{H}_{I,\mathrm{eff}}^{(1)}(\bm{\tau}) \hat{U}_{0}(\bm{\tau})
\end{equation}
such that the factorization $\hat{U}_{\mathrm{eff}}(\bm{\tau};\lambda) = \hat{U}_{0}^{\prime}(\bm{\tau}) \hat{U}_{I}^{\prime}(\bm{\tau};\lambda)$ returns the  new interaction picture equation
\begin{equation}\label{eq:qat_multiplescales:multitime_averaged_system_standard_form}
i\left(\frac{1}{\lambda} \frac{d}{ds}\right)\hat{U}^{\prime}_I(\bm{\tau};\lambda) = \sum_{n=2}^{\infty} \lambda^{n-1} \, \hat{H}_{I,\mathrm{eff}}^{\prime \, (n)}(\bm{\tau}) \hat{U}^{\prime}_I(\bm{\tau};\lambda) 
\end{equation}
where 
\begin{equation}
    \hat{H}_{I,\mathrm{eff}}^{\prime \, (n)}(\bm{\tau}) = \hat{U}_{0}^{\prime \, \dagger} (\bm{\tau}) \hat{H}_{I,\mathrm{eff}}^{(n)} (\bm{\tau}) \hat{U}_{0}^{\prime}(\bm{\tau}).
\end{equation}
Applying the PETS technique, with the identification $(s,\tau_1) \rightarrow (\tau_1,\tau_2)$ in the new interaction picture, enables the separation of slow $\tau_1$-time dynamics from even slower $\tau_2$-time dynamics.
By iterating this procedure, PETS can be used to successively isolate increasingly slower timescale dynamics.

In the Multiple PETS renormalization procedure, a timescale renormalization step is represented as
\begin{equation}\label{eq:qat_multiplescales:multitime_renormalization_step}
    \hat{U}_I(\sigma,\bm{\tau}) \overset{\hat{U}_{\mathrm{ren}}^{\prime}}{\Longrightarrow} \; \hat{U}^{\prime}_I(\bm{\tau}),
\end{equation}
which describes applying the PETS technique followed by the transformation into eq.~\eqref{eq:qat_multiplescales:multitime_averaged_system_standard_form}.
Explicitly, the step corresponds to the factorization
\begin{equation}\label{eq:qat_multiplescales:multitime_renormalization_factorization}
    \hat{U}_I(\sigma,\bm{\tau};\lambda) = \hat{U}_{\mathrm{ren}}^{\prime}(\sigma,\bm{\tau};\lambda) \hat{U}^{\prime}_I(\bm{\tau};\lambda)
\end{equation}
where 
\begin{equation}
    % \hat{U}_{\mathrm{ren}}^{\prime}(\sigma,\bm{\tau};\lambda) = \exp(-i \hat{\Phi}(\sigma,\bm{\tau};\lambda)) \hat{U}_{0}^{\prime}(\bm{\tau})
        \hat{U}_{\mathrm{ren}}^{\prime}(\sigma,\bm{\tau};\lambda) = \hat{U}_\mathrm{fast}(\sigma,\bm{\tau};\lambda) \hat{U}_{0}^{\prime}(\bm{\tau})
\end{equation}
with $ \hat{U}_\mathrm{fast}(\sigma,\bm{\tau};\lambda) = \exp(-i\hat{\Phi}(\sigma,\bm{\tau};\lambda))$ generated with the PETS approach.
%and $\hat{H}_{I,\mathrm{eff}}(\bm{\tau};\lambda)$ 
%asymptotically generated by solving \eqref{eq:qat_multiplescales:multi_QAT_homological} with the regularization condition in \eqref{eq:qat_multiplescales:multi_QAT_regularization}.
For example, in the simplest case where $\hat{H}_I(s) \equiv \hat{H}_I(\sigma)$ depends only on the fast timescale $\sigma$, the renormalization step describes applying the far-detuned QAT of Sec.~\eqref{sec:qat} to generate a time-independent effective Hamiltonian, which is then transformed into a time-dependent perturbation problem.
We remind the reader that the explicit reference to multiple timescales serves as a visual aid.
The PETS approach allows us to avoid the cumbersome frequency decomposition with the relevant timescales naturally emerging during each renormalization step.

% Similarly, for $\hat{H}_I(s) \equiv \hat{H}_I(\sigma, \tau)$ for a single slow timescale, the renormalization step simply describes the procedure in appendix~\eqref{appendix:effective_equation_PT}.
The Multiple PETS renormalization procedure is summarized as the following iterative sequence:
\begin{equation}\label{eq:qat_multiplescales:qat_multiplescales_renormalization}
    \begin{aligned}
    \hat{U}_I(\sigma,\bm{\tau}) &\overset{\hat{U}_{\mathrm{ren}}^{\prime}}{\Longrightarrow} \; \hat{U}^{\prime}_I(\bm{\tau}) \; \overset{\hat{U}_{\mathrm{ren}}^{\prime\prime}}{\Longrightarrow} \;
    \hat{U}^{\prime\prime}_I(\bm{\tau}\setminus\{\tau_1\}) \\
    &\quad\Longrightarrow \; \ldots \;
    \overset{\hat{U}_{\mathrm{ren}}^{\prime\cdots\prime}}{\Longrightarrow} \; \hat{U}^{\prime\cdots\prime}_I(\tau_M).
    \end{aligned}
\end{equation}
This procedure is entirely equivalent to a multi-timescale derivative expansion, where each renormalization step (denoted by the primes) introduces a new interaction picture by performing a partial average over the fastest timescale.
It is important to note that each renormalization step requires the corresponding `unperturbed' system to be explicitly solvable within the perturbation problem.
If explicit solvability is not possible, the procedure terminates at that step, and the resulting effective Hamiltonian must be resolved numerically.

\section{Conclusion}\label{sec:conclusion}
In this work, we introduced Quantum Averaging Theory (QAT) as a generalized unitarity-preserving analytic framework for analyzing periodically and almost-periodically driven quantum systems across multiple timescales. 
By integrating the Magnus expansion \cite{magnusExponentialSolutionDifferential1954} with the method of averaging on multiple timescales \cite{frascaTheoryQuantumResonance1998,frascaUnifiedTheoryQuantum1992,nayfehMethodMultipleScales2000}, QAT provides a systematic perturbative approach to describe both far-detuned (high-frequency, off-resonant) and near-resonant interactions, bridging the gap between existing analytic approaches. 
As in prior quantum averaging frameworks \cite{rahavEffectiveHamiltoniansPeriodically2003,goldmanPeriodicallyDrivenQuantum2014,eckardtHighfrequencyApproximationPeriodically2015,bukovUniversalHighFrequencyBehavior2015,casasFloquetTheoryExponential2001,kuwaharaFloquetMagnusTheory2016}, QAT generates an effective Hamiltonian description for the slow-time evolution while retaining fast oscillatory effects within a separate dynamical phase operator. 
However, by allowing the effective Hamiltonian to capture slowly-varying interactions this approach generalizes Floquet-based methods while refining and extending standard approximation techniques, such as the rotating-wave approximation (RWA). 
We demonstrated that, in both detuning regimes, the QAT expansion rapidly converges toward exact numerical solutions, underscoring its high accuracy. 
For two touchstone problems in quantum optics, we showed that the QAT expansion can be mapped onto a closed-form solution.

Compared to conventional time-dependent perturbative methods, QAT provides several key advantages. 
First, it preserves unitarity at all orders, eliminating non-Hermitian artifacts commonly introduced by truncations in non-unitary expansions. 
Second, it enables high-order corrections in a controlled manner, systematically improving upon approximations such as RWA and adiabatic elimination. Third, unlike Floquet or Van Vleck expansions \cite{casasFloquetTheoryExponential2001, eckardtHighfrequencyApproximationPeriodically2015}, which rely on strict periodicity and off-resonant driving, QAT is applicable to single- and multi-frequency systems with near- and off-resonant driving, making it a versatile tool for a wide range of driven quantum systems. 
By explicitly capturing multi-scale interactions, QAT remains effective even when disparate frequency scales and near-resonant interactions are present—regimes where standard high-frequency expansions break down.

Our results illustrate that high-frequency effects are both unavoidable and significant, even in weakly driven systems. 
By capturing multi-timescale effects, QAT can aid in analyzing complex driven systems, providing clearer insight into phenomena that were previously hard to model. 
Taking these effects into account is particularly relevant for error-resilient gate designs in quantum computing and tailored driving schemes in analog quantum simulations. 
To demonstrate this capability, in the companion paper \cite{barajasMultiTimescaleCoherentControl2025} we analyzed entangling gate performance under the well-known M{\o}lmer-S{\o}rensen interaction, showing how the high-frequency corrections can be suppressed to improve gate fidelity. 
While the present formulation applies to closed, unitary dynamics, a more realistic description must account for dissipation and decoherence, which inevitably arise in physical systems. 
An important direction for future work is extending QAT to open quantum systems where incorporating environmental interactions within a master equation framework would significantly broaden its applicability. 
Moreover, investigating the range of validity for QAT in both strongly driven and many-body quantum systems remains an open question for further research.

\begin{acknowledgments}
    We gratefully acknowledge the late Alex Levine for his early guidance and inspiring discussions.
    We also thank Robijn Bruinsma for valuable comments on the manuscript.
    This work was supported by the National Science Foundation under Grants PHY-2207985 and OMA-2016245, and by the U.S. Army Research Office under Grant W911NF-24-S-0004.
\end{acknowledgments}

\subsection*{CRediT Author Statement}

\textbf{Kristian D. Barajas:} Conceptualization, Investigation, Formal analysis, Writing – Original Draft, Writing – Review \& Editing 

\textbf{Wesley C. Campbell:} Supervision, Validation, Writing – Review \& Editing, Funding Acquisition

\begin{appendices}

\section{Exact, Non-Linear Differential Equation for the Dynamical Phase }\label{appendix:Magnus_type_expansions}
By inserting \eqref{eq:qat:interaction_propagator_factorized} into \eqref{eq:qat:interaction_eq} and transforming into the effective interaction picture we find an exact expression for $\hat{H}_{I,\mathrm{eff}}$ depending only on $\hat{H}_I$ and $\hat{\Phi}$ given by
\begin{equation}\label{eq:appendix:effHamiltonian_eq}
	\begin{aligned}
		\hat{H}_{I,\mathrm{eff}}(s) &= \hat{U}_\mathrm{fast}^\dagger(s) \, \hat{H}_I(s) \, \hat{U}_\mathrm{fast}(s) - i \, \hat{U}_\mathrm{fast}^\dagger(s) \, \partial_s \hat{U}_\mathrm{fast}(s) \\  
           &= e^{i \hat{\Phi}(s)} \hat{H}_I(s) e^{-i \hat{\Phi}(s)} - i \, e^{i \hat{\Phi}(s)} \frac{\partial}{\partial s} \left( e^{-i \hat{\Phi}(s)} \right),
	\end{aligned}
\end{equation}
where the last term on line two requires care since, in general, $[\hat{\Phi},\partial_{s}\hat{\Phi}]$ do not commute. 
To ensure proper operator ordering we proceed with a Lie theoretic approach proposed by Magnus \cite{magnusExponentialSolutionDifferential1954}.

We remark that the following derivation draws on two important features of the exponential map: (i) $\hat{\Phi}$ preserves the Lie algebra $\mathfrak{g}$ generated by $\hat{H}_I$, restricting the map $\exp:\mathfrak{g} \mapsto G$ to the Lie group structure $G$ for the set of group-preserving operations, and (ii) the exponential map has the easily computed inverse property $\exp(i \hat{\Phi})^{-1} = \exp(-i \hat{\Phi})$.
Equipped with the matrix Lie bracket $[ \hat{A},\hat{B}] = \hat{A} \hat{B} - \hat{B} \hat{A}$ and the adjoint action $\mathrm{ad}_{\hat{A}}: \mathfrak{g} \mapsto \mathfrak{g}$ where
\begin{equation}\label{eq:appendix:adjoint_action}
    \mathrm{ad}_{\hat{A}}^{k} (\hat{B}) \doteq %[(\hat{A}) ^{k},\hat{B}]=
	\begin{cases}
		\hat{B} & \text{if } {k=0} \\
		[ \hat{A} , \hat{B} ] & \text{if } {k = 1} \\
		[\hat{A}, \mathrm{ad}_{\hat{A}}^{k-1}(\hat{B})] & \text{otherwise}
    \end{cases}
\end{equation}
for all $k \in \mathbb{N}$ and $\hat{A},\hat{B} \in \mathfrak{g}$, \eqref{eq:appendix:effHamiltonian_eq} can be expressed in terms of Lie group-theoretic operators.
From the Baker-Campbell-Hausdorff identity we have
\begin{equation}\label{eq:appendix:MA_interaction_BCH}
\begin{aligned}
\hat{U}_\mathrm{fast}^{\dagger} \, \hat{H}_I \,  \hat{U}_\mathrm{fast} &= \exp(\mathrm{ad}_{i \hat{\Phi}}) \hat{H}_I \\
&= \sum_{k=0}^{\infty} \frac{1}{k!} \mathrm{ad}^k_{i\hat{\Phi}}(\hat{H}_I)
\end{aligned}
\end{equation}
and from the differentiation of the exponential map
\begin{equation}\label{eq:appendix:dexp_op}
    \begin{aligned}
    \mathrm{dexp}_{i \hat{\Phi}} \left(\partial_{s} \hat{\Phi} \right) &= i \, \hat{U}_\mathrm{fast}^{\dagger} \partial_{s} \hat{U}_\mathrm{fast} \\
    &= \frac{\exp(\mathrm{ad}_{i \hat{\Phi}})-I}{\mathrm{ad}_{i\hat{\Phi}}} \left(\partial_s \hat{\Phi} \right)\\
    &= \sum_{k=0}^{\infty} \frac{1}{(k+1)!} \mathrm{ad}^{k}_{i\hat{\Phi}} \left(\partial_s \hat{\Phi} \right),
    \end{aligned}
\end{equation}
which enforces proper operator ordering.
Rearranging terms in \eqref{eq:appendix:effHamiltonian_eq} and assuming invertability of the $\mathrm{dexp}_{i \hat{\Phi}}$ operator we have
\begin{equation}\label{eq:appendix:exponent_eqn_unsimplified}
    \partial_s \hat{\Phi} 
    = \frac{\hat{X}}{e^{\hat{X}} - I}  \left(e^{\hat{X}} \hat{H}_{I} - \hat{H}_{I,\mathrm{eff}} \right), \quad \hat{X} \equiv \mathrm{ad}_{i \hat{\Phi}}
\end{equation}
where the inverse $\mathrm{dexp}_{i \hat{\Phi}}$ operator is formally given by
\begin{equation}\label{eq:appendix:inverse_dexp}
\mathrm{dexp}^{-1}_{i \hat{\Phi}} = \frac{\hat{X}}{e^{\hat{X}} - I} \doteq \sum_{k=0}^{\infty} \frac{B_k}{k!} \mathrm{ad}^{k}_{i\hat{\Phi}} ( \, \cdot \, )
\end{equation}
where $B_k$ are the Bernoulli numbers. 
Recognizing that $\hat{X} e^{\hat{X}} / (e^{\hat{X}} - I) = -\hat{X} / (e^{-\hat{X}}-I)$ and $\pm \hat{X} \equiv \mathrm{ad}_{\pm i \hat{\Phi}}$, eq.~\eqref{eq:appendix:exponent_eqn_unsimplified} is expressed more conveniently as
\begin{equation}\label{eq:appendix:magnus_exponent_eqn}
\begin{aligned}
     \partial_s \hat{\Phi} &= \hat{H}_{\mathrm{ME}} - \hat{H}_{\mathrm{reg}} \\ 
     &= \sum_{k\ge0} \frac{B_k}{k!} \mathrm{ad}_{i \hat{\Phi}}^{(k)} \left(  (-1)^k \, \hat{H}_{I} - \hat{H}_{I,\mathrm{eff}} \right)
\end{aligned}
\end{equation}
where $\hat{H}_{\mathrm{ME}} = \mathrm{dexp}^{-1}_{-i \hat{\Phi}}(\hat{H}_{I})$ is the generator of the standard Magnus expansion and $\hat{H}_{\mathrm{reg}} = \mathrm{dexp}^{-1}_{i \hat{\Phi}}(\hat{H}_{I,\mathrm{eff}}) $ is a degree of freedom to be leveraged \cite{blanesMagnusExpansionIts2009,schererNewPerturbationAlgorithms1997}.
While the expression is exact, in practice it can only be solved asymptotically in a Magnus-type expansion where $\hat{H}_{I,\mathrm{eff}}=0$ yields the standard Magnus expansion \cite{magnusExponentialSolutionDifferential1954,blanesMagnusExpansionIts2009}.
Finally, we draw attention to a subtle point inferred from \eqref{eq:appendix:magnus_exponent_eqn} and the adjoint action in \eqref{eq:appendix:adjoint_action}: for $\hat{H}_I \in \mathfrak{g}$, we must have $\hat{\Phi},\hat{H}_{I,\mathrm{eff}} \in \mathfrak{g}$ ensuring that $\exp(\pm i  \hat{\Phi}) \in G$ where $G$ is the group of symmetry-preserving transformations as required. 
Hence, \eqref{eq:appendix:magnus_exponent_eqn} preserves the Lie group structure to all finite orders of approximation.

%%%%%%%%%%%%%%%%%%%%%%%%%%%%%%%%%%

\section{Non-Uniqueness of Floquet Solution in Far-Detuned Rabi Problem}\label{appendix:Rabi_problem}

By closing the summation of the asymptotic series with our dynamical phase ansatz, we find in the highly-detuned limit that quantum averaging theory provides an explicit solution to the interaction picture Rabi Hamiltonian that obeys Floquet theorem.
Perhaps a more familiar solution method follows Rabi's original proposal to use a $\hat{z}$-rotation propagator $\hat{U}_r(t) = \exp{(i \, \Delta \, \hat{\sigma}_z \, t/2)}$ to transform $\hat{H}_I(t,\lambda)$ into a constant Hamiltonian $\hat{H}_r = \frac{\Delta}{2} \sigma_z + \frac{\Omega}{2} \sigma_x$, which is then solved by matrix exponentiation. 
This approach may be related back to the QAT interaction propagator up to a constant gauge transformation, demonstrating the nonuniqueness property of Floquet's theorem.
More explicitly, in real time $t = s/\omega_{eg}$ with frequency $\omega \equiv \Lambda_\omega \, \omega_{eg}$, we have the relation
\begin{equation}
	\begin{aligned}\label{eq:ex:Rabi_interaction_propagator_relation}
	\hat{U}_I(t,t_0=0) ={}&  \hat{U}_r(t) e^{-i \, \left(\Delta \, \hat{\sigma}_z + \Omega \, \hat{\sigma}_x \right) t/2 } \\
      ={}& \hat{U}_r(t) {\hat{R}(\theta/2)} e^{-i \,\tilde{\Omega} \hat{\sigma}_z t/2} \, \hat{R}^{\dagger}(\theta/2) \\
	={}& \hat{U}_{\mathrm{fast}}(t) \hat{U}_{\mathrm{eff}}(t) \hat{R}^{\dagger}(\theta/2)
\end{aligned}
\end{equation}
where
\begin{equation}
\begin{aligned}
    \hat{U}_{\mathrm{fast}}(t) \equiv{}& \hat{U}_r(t) \hat{R}(\theta/2) \hat{U}_r^\dagger (t) \\
    \hat{U}_{\mathrm{eff}}(t) \equiv{}& \hat{U}_r(t) \exp(-i \, \tilde{\Omega}\hat{\sigma}_z t/2)
\end{aligned}
\end{equation}
with the rotation matrix
\begin{equation}
    \hat{R}(\theta/2) = \begin{pmatrix}
			\cos{(\theta/2)} & -\sin{(\theta/2)}  \\
			\sin{(\theta/2)} & \cos{(\theta/2)} 
    \end{pmatrix}
\end{equation}
and the generalized Rabi frequency $\tilde{\Omega} = \sqrt{\Delta ^2 + \Omega ^2 }$.

\section{Approximate Dynamics for The Time-Dependent Effective Equation}\label{appendix:effective_equation_PT}

Suppose no explicit solution exists for the truncated effective interaction picture equation and an analytic approximation is desired.
From the two-timescale derivative expansion, we find that the effective interaction picture equation truncated at some finite order $N \ge 1$ can be equivalently expressed as
\begin{equation}\label{eq:appendix:twotime_average_eq}
    i \lambda \, \partial_\tau \hat{U}_\mathrm{eff}^{[N]}(\tau,\lambda) = \hat{H}_{I,\mathrm{eff}}^{[N]}(\tau, \lambda) \, \hat{U}_\mathrm{eff}^{[N]}(\tau,\lambda)
\end{equation}
where we've identified $\partial_s \rightarrow \lambda \partial_\tau$.
The $\tau$-dependent form demonstrates that even systems with a single timescale as studied in \cref{sec:qat} invariably exhibit two timescale dynamics.
Expanding $\hat{H}_{I,\mathrm{eff}}(\tau, \lambda)$ in powers of $\lambda$ reveals the slow $\tau$-time perturbation problem
\begin{equation}\label{eq:appendix:effective_equation_PT:twotime_perturbation_problem}
    i \partial_\tau \hat{U}_\mathrm{eff}^{[N]} = \Bigl( \hat{H}_{I,\mathrm{eff}}^{(1)} + \sum_{n=1}^{N-1} \lambda^n \hat{H}_{I,\mathrm{eff}}^{(n+1)} \Bigr) \hat{U}_\mathrm{eff}^{[N]}
\end{equation}
where the explicit $\tau$-time dependence is omitted.
The form strongly suggests that the adiabatic dynamics are predominately governed by the first-order approximation $\hat{H}_{I,\mathrm{eff}}^{(1)}(\tau)$ resulting from the rotating-wave approximation. 

If the time-evolution of the first-order approximation is explicitly solvable, then the effective equation can be expressed in an interaction picture with respect to $\hat{H}_{I,\mathrm{eff}}^{(1)}(\tau)$ such that the slow-time perturbation problem is compatible with exponential perturbation theory, allowing further analytic treatment.
Under these assumptions, let the first-order approximation define the ``unperturbed'' system, i.e.
\begin{equation}\label{eq:appendix:effective_equation_PT:twotime_perturbation_problem_o1}
    i\partial_{\tau} \hat{U}_{\mathrm{eff},0}^{\prime}(\tau) = \hat{H}_{I,\mathrm{eff}}^{(1)}(\tau) \hat{U}_{\mathrm{eff},0}^{\prime}(\tau).
\end{equation}
where $\hat{U}_{\mathrm{eff},0}^{\prime}(\tau)$ is explicitly solvable.
Inserting the unitary transformation $\hat{U}_{\mathrm{eff}}^{[N]}(\tau,\lambda) = \hat{U}_{\mathrm{eff},0}^{\prime}(\tau) \hat{U}^{\prime [N-1]}_{\mathrm{eff}}(\tau, \lambda)$ into \eqref{eq:appendix:effective_equation_PT:twotime_perturbation_problem} yields the new slow-time interaction picture
\begin{equation}\label{eq:appendix:effective_equation_PT:twotime_perturbation_problem_standard_form}
    i \partial_\tau \hat{U}_{\mathrm{eff}}^{\prime \, [N-1]}(\tau,\lambda) = \sum_{n=1}^{N-1} \lambda^{n} \hat{H}_{I,\mathrm{eff}}^{\prime \, (n+1)}(\tau) \hat{U}_{I,\mathrm{eff}}^{\prime \, [N-1]}(\tau,\lambda)
\end{equation}
where 
\begin{equation}
    \hat{H}_{I,\mathrm{eff}}^{\prime \, (n)}(\tau) = \hat{U}_{\mathrm{eff},0}^{\prime \, \dagger}(\tau)  \hat{H}_{I,\mathrm{eff}}^{(n)} (\tau) \hat{U}_{\mathrm{eff},0}^{\prime}(\tau).
\end{equation}
The new truncation order reflects that \eqref{eq:appendix:effective_equation_PT:twotime_perturbation_problem_standard_form} need only be analyzed with exponential perturbation theory up to $\mathcal{O}(\lambda^{N-1})$ such that 
\begin{equation}\label{eq:appendix:effective_equation_PT:twotime_perturbation_problem_solution}
    \hat{U}_{\mathrm{eff}}^{[N]} (\tau,\lambda) = \hat{U}_{\mathrm{eff},0}^{\prime}(\tau) \hat{U}_{\mathrm{eff}}^{\prime \, [N-1]}(\tau,\lambda) + \mathcal{O}(\lambda^{N})
\end{equation}
as desired. 
Inserting into \eqref{eq:qat:QAT_interaction_approx} completes the process.

\section{Multi-Timescale Derivative Expansion Procedure}\label{appendix:multi_timescale_expansion}
As in the two-timescale QAT approach, using the chain-rule we introduce the multi-timescale derivative expansion
\begin{equation}\label{eq:qat_multiplescales:qat_multiplescales_derivative}
	\frac{d}{d s} \doteq \left(\frac{\partial}{\partial \sigma}\right)_{\bm{\tau}} + \sum_{j=1}^{M} \lambda^j  \left(\frac{\partial}{\partial \tau_j}\right)_{\sigma, \bm{\tau}\setminus\{\tau_j\}}
\end{equation}
with  $M+1$ independent timescale variables.
As before, integration over a specified timescale requires holding all others constant.
To accommodate multiple timescales we proceed as in \cref{sec:qat_multiplescales:twotime} and replace the time-derivative in \eqref{eq:qat:QAT_homological} to yield the multi-timescale homological equation
\begin{multline}\label{eq:qat_multiplescales:multi_QAT_homological}
    \partial_\sigma \hat{\Phi}^{(n)}(\sigma,\bm{\tau}) = \hat{\mathcal{H}}^{(n)}_{\Phi}(\sigma,\bm{\tau}) - \hat{H}_{I,\mathrm{eff}}^{(n)}(\bm{\tau}) \\
    - \sum_{j=1}^{\mathclap{\min(M,n)}}\partial_{\tau_{j}} \hat{\Phi}^{(n-j)}(\sigma,\bm{\tau})
\end{multline}
with $\hat{\mathcal{H}}_{\hat{\Phi}}^{(n)}$ defined in \eqref{eq:qat:QAT_aux_generator}.
The solution to the homological equation is regularized by the familiar conditions
\begin{subequations}\label{eq:qat_multiplescales:multi_QAT_regularization}
\begin{align}
    \timeavg{\hat{\Phi}^{(n)}(\sigma,\bm{\tau})}{\sigma} ={}& 0 \\
    \hat{H}^{(n)}_{I,\mathrm{eff}}(\bm{\tau}) ={}& \timeavg{\hat{\mathcal{H}}^{(n)}_{\hat{\Phi}}(\sigma,\bm{\tau})}{\sigma}
\end{align}
\end{subequations}
with the time-averaging procedure for multiple scales
\begin{equation}\label{eq:qat_multiplescales:time_average_twotime}
	\timeavg{\hat{A}}{\sigma}(\bm{\tau}) \doteq \lim_{T \to \infty} \frac{1}{T} \int_{{0}}^{{T}} {\hat{A}(\sigma,\bm{\tau})} \: d{\sigma}
\end{equation}
which only averages over the $\sigma$-time effects.
The $\sigma$-averaged effective system now depends on $\bm{\tau}$-times and is given by
\begin{equation}\label{eq:qat_multiplescales:multitime_average_eq}
i \frac{d}{ds}\hat{U}_{\mathrm{eff}}(\bm{\tau},\lambda) = \hat{H}_{I,\mathrm{eff}}(\bm{\tau},\lambda) \hat{U}_{\mathrm{eff}}(\bm{\tau},\lambda)
\end{equation}
where $d/ds$ is the multi-timescale derivative in \eqref{eq:qat_multiplescales:qat_multiplescales_derivative}.
Rewriting \eqref{eq:qat:QAT_eff_system_slowtime} as a $\bm{\tau}$-time perturbation problem we have
\begin{equation}%\label{eq:qat_multiplescales:multitime_perturbation_problem}
\begin{split}
    i \! \left(\frac{1}{\lambda} \frac{d}{ds}\right)\hat{U}_{\mathrm{eff}}(\bm{\tau},\lambda) ={}& \Bigl( \hat{H}_{I,\mathrm{eff}}^{(1)}(\bm{\tau}) \; + \\
    & \! \sum_{n=2}^{\infty} \lambda^{n-1} \hat{H}_{I,\mathrm{eff}}^{(n)}(\bm{\tau}) \Bigr) \hat{U}_{\mathrm{eff}}(\bm{\tau},\lambda)
\end{split}
\end{equation}
where the rescaled time-derivative $\lambda^{-1} d/ds$ should be interpreted as $\partial_{\tau_1} + \sum_{j=2}^{M} \lambda^{j-1} \partial_{\tau_j}$ (dropping $\partial_\sigma$ since $\hat{U}_I(\bm{\tau},\lambda)$ doesn't depend on $\sigma$). 

\end{appendices}

\bibliography{2024_QAT_PRA}

\end{document}